# Trion-mediated quantum operations on a double-dot charge qubit

A.V. Tsukanov

**NRC "Kurchatov institute", Moscow, Russia**

**Email Address:** a-v-ts@mail.ru

**Abstract:** A scheme for controlling a charge qubit on a semiconductor asymmetric double quantum dot with suppressed tunnel coupling between individual quantum dots is proposed. Laser pulses convert the logical single-electron states of the qubit into auxiliary trion states, for which the resonant tunneling condition is restored due to the compensation of structural asymmetry by the Coulomb interaction of the particles. The conditions for performing orthogonal single-qubit rotations are formulated. For a two-dimensional structure, the parameters of the qubit are calculated, and the dependence of the fidelity of the inversion operation on the parameters of the laser field and the rates of dissipative processes is obtained. It is shown that the considered algorithm is characterized by high speed and reliability.

## 1. Introduction

Solid-state semiconductor nanostructures with optical control attract the attention of researchers with the possibility of creating multifunctional quantum devices based on them, consisting of a large number of identical elements (see reviews [1 – 3]). In particular, quantum bits on quantum dots (QD) and methods of their control are the subject of intensive study (see reviews [4 – 12]). There are several options for encoding quantum information into electronic (see reviews [4, 5]), spin (see reviews [6, 7]), and exciton (see reviews [8, 9]) states of QDs. Quantum operations are performed using external electromagnetic fields that induce resonant optical and/or tunneling transitions between the energy levels of single QDs and their complexes (so-called "artificial molecules"). For qubits on exciton states, laser pulses and microresonators with a frequency equal to the frequency of the transition from the vacuum state of the QD to the exciton state are used. Changing the state of a qubit on spatially separated single-electron levels of a double quantum dot (DQD), known as charge qubit, is performed by electric field pulses of the gates [13, 14]. To date, basic operations on these qubits have already been demonstrated [15 – 20].

Each of these encoding and control schemes has its own advantages and disadvantages. For example, exciton qubit operations are highly accurate and fast, but the exciton lifetime in a QD is less than 1 ns, making it unsuitable for storing quantum information. On the other hand, charge qubit based on spatially separated DQD ground states is resistant to relaxation. However, a large number of metal gate contacts generate electrical noise, leading to dephasing of the DQD's electronic state [21], and the qubits themselves interact in an uncontrolled manner [22].

In our work, we will consider the principle of controlling the charge qubit, which, while preserving its useful properties, allows us to implement the required evolution of the state vector using fully optical means, similar to the exciton qubit. This concept is based on the addition of a new element, the auxiliary trion state, which is a three-particle complex consisting of two electrons and a hole [23 – 33]. The idea of using a trion to rapidly switch on/off the tunneling within DQD was previously proposed in Ref. [29]. The authors of Ref. [30] demonstrated a spin qubit where the trion functions as one of the logical states. A theoretical model with encoding of quantum information into spin states of an electron located in one of the QDs, where the qubit control is carried out by spin-charge conversion, was proposed in the works [34 – 37]. Here, the trion plays the role of an intermediate element in the Raman scheme, transforming the spin state of the electron into one of the spatially separated orbital states of the DQD. In addition, the trion can also be used for the optical measurement of the charge qubit [38].

In our proposed scheme, the trion is formed by a laser pulse that generates an exciton in one of the charge qubit's QDs. The additional energy of the Coulomb interaction compensates for the energy difference between the logical states of the qubit, which is caused by the structural asymmetry of the DQD, transitioning it from the tunnel blockade regime to the resonant tunneling regime. Thus, the trion acts as an intermediate state that performs the required single-qubit rotation, after which a second laser pulse annihilates the exciton, returning the transformed state to the logical subspace of the DQD. A similar algorithm exists for the phase shift operation. It is shown that the fidelity of single-qubit operations can be quite high ($F > 0.95$) for structures that have already been created. Moreover, the trion can be used to entangle the states of two or more qubits in schemes with additional optical modes in a waveguide or microresonator. Section 2 describes a model of optical transitions in the DQD between the electronic and trion subspaces. Single-qubit operations in the coherent approximation and their implementation conditions are discussed in Section 3. In Section 4, an example is provided for calculating the parameters of a qubit formed by two-dimensional GaAs/InGaAs QDs with a Gaussian confinement potential and lateral tunneling coupling. Section 5 focuses on modeling a quantum inversion operation, taking into account dissipative effects, and calculating its accuracy when varying the parameters of the laser field. The results are discussed and compared with other models in Section 6.

## 2. Optical and tunnel transitions in asymmetric DQD involving trion states

Consider a DQD formed by crystalline QD A and QD B based on the GaAs/$In_xGa_{1-x}As$ compound, where the regions with indium content correspond to potential wells for electrons and holes (Fig. 1). We assume that the DQD spectrum contains single-electron and negatively charged trion (2e+1h) states. Let us represent the structure Hamiltonian as a sum of one- and two-particle Hamiltonians,

$$H_{DQD} = H\left(\mathbf{r}_{e,1}\right) + H\left(\mathbf{r}_{e,2}\right) + H\left(\mathbf{r}_{h}\right) + 2/\left|\mathbf{r}_{e,1} - \mathbf{r}_{e,2}\right| - 2/\left|\mathbf{r}_{e,1} - \mathbf{r}_{h}\right| - 2/\left|\mathbf{r}_{e,2} - \mathbf{r}_{h}\right|, \quad (1)$$

where $\mathbf{r}_{e,1}$, $\mathbf{r}_{e,2}$, and $\mathbf{r}_h$ are the electron and hole radius-vectors. Hereinafter, all parameters are expressed in relative atomic units [8], $Ry^* = Ry\left(m_e^*/m_e\varepsilon^2\right)$ and $a_B^* = a_B\left(m_e\varepsilon/m_e^*\right)$, where $Ry$ = 13.6 eV is the Rydberg energy, $a_B = 0.5292\times10^{-10}$ $m$ is the Bohr radius, $m_e$ is the mass of the electron, $m_e^* = 0.067m_e$ is the effective mass of the electron (GaAs), and $\varepsilon = 12.5$ is the dielectric constant (GaAs). In each of the QD, there are single-particle states with wave functions $\tilde{\varphi}_i\left(\mathbf{r}_h\right)$ for the hole and $\varphi_i\left(\mathbf{r}_e\right)$ for the electron with energies $\tilde{\varepsilon}_i$ and $\varepsilon_i$ ($i$ = A, B), respectively. Let us consider them to be orthonormal, i.e., $\int d\mathbf{r}_e \varphi_i^*\left(\mathbf{r}_e\right)\varphi_j\left(\mathbf{r}_e\right) = \delta_{ij}$ and $\int d\mathbf{r}_h \tilde{\varphi}_i^*\left(\mathbf{r}_h\right)\tilde{\varphi}_j\left(\mathbf{r}_h\right) = \delta_{ij}$.

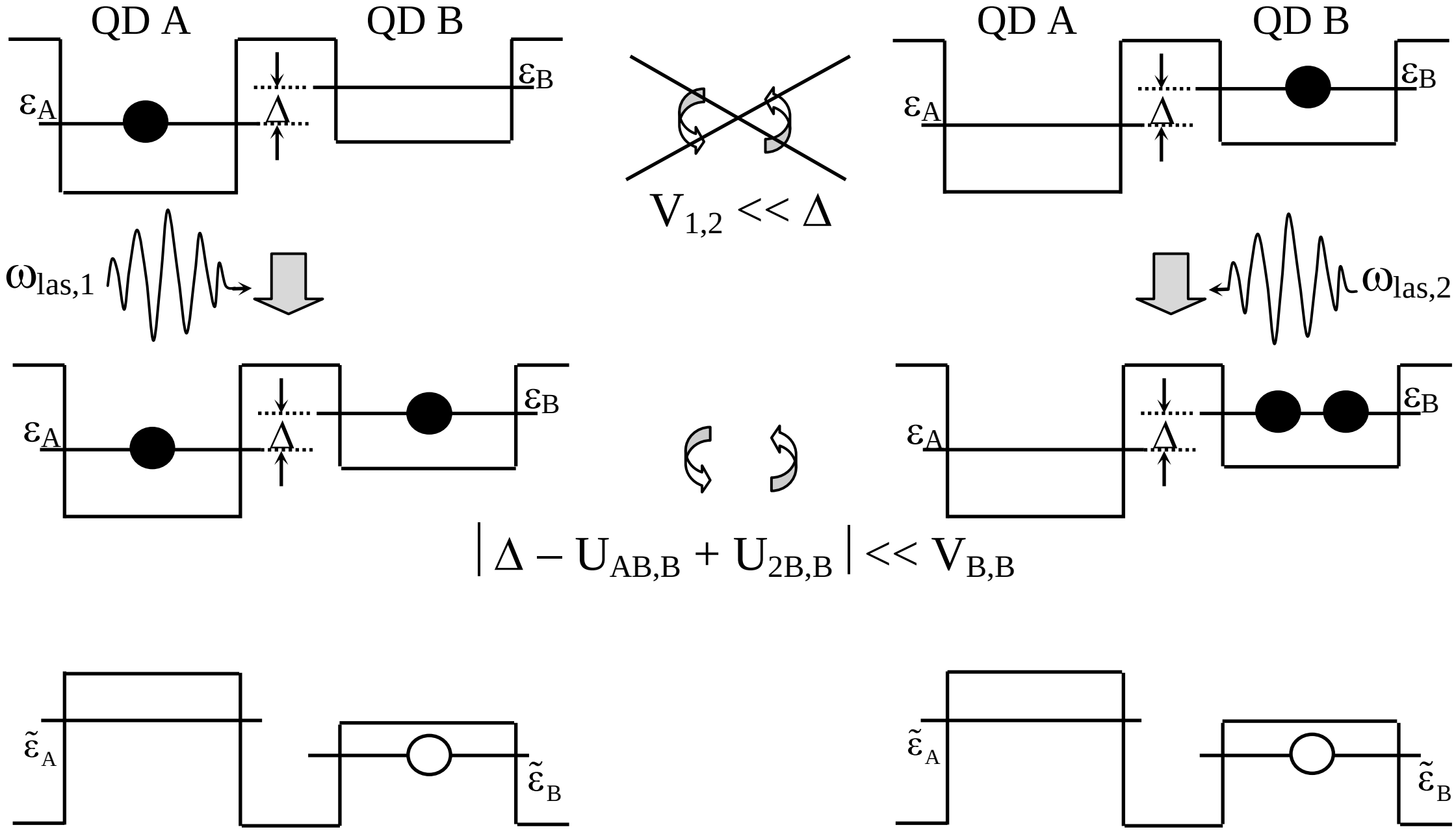

*Fig. 1. Energy diagram of a charge qubit on a single-electron asymmetric DQD with suppressed tunneling between logical states (top). Stimulation of resonant tunneling electron transfer between "dressed" (trion) DQD states using a bichromatic laser pulse that generates an exciton in QD B (bottom). The electron and hole are represented by black and white circles, respectively.*

As three-particle basis states, we choose the symmetrized combinations of products of one-particle orbitals corresponding to spin singlets and triplets. We use the conventional notation for trions, ${}^{n_A,n_B}_{p_A,p_B}X_S^{-1}$, which reflects the distribution of particles over the QD states, where $0 \le n_i$ $(p_i) \le 2$ is the number of electrons (holes) occupying the state in the $i$-th QD, and $S = s$, $t$ indicates the symmetry of the spin component (singlet or triplet). We divide them into two groups, one of which has a hole localized in QD A,

$$
\begin{aligned}
{}_{1,0}^{2,0}X_{s}^{-1} &= \tilde{\varphi}_A(\mathbf{r}_h)\varphi_A(\mathbf{r}_{e,1})\varphi_A(\mathbf{r}_{e,2})\otimes\chi_s,\\
{}_{1,0}^{0,2}X_{s}^{-1} &= \tilde{\varphi}_A(\mathbf{r}_h)\varphi_B(\mathbf{r}_{e,1})\varphi_B(\mathbf{r}_{e,2})\otimes\chi_s,\\
{}_{1,0}^{1,1}X_{s}^{-1} &= \tilde{\varphi}_A(\mathbf{r}_h)\frac{\varphi_A(\mathbf{r}_{e,1})\varphi_B(\mathbf{r}_{e,2})+\varphi_A(\mathbf{r}_{e,2})\varphi_B(\mathbf{r}_{e,1})}{\sqrt{2}}\otimes\chi_s,\\
{}_{1,0}^{1,1}X_{t,k}^{-1} &= \tilde{\varphi}_A(\mathbf{r}_h)\frac{\varphi_A(\mathbf{r}_{e,1})\varphi_B(\mathbf{r}_{e,2})-\varphi_A(\mathbf{r}_{e,2})\varphi_B(\mathbf{r}_{e,1})}{\sqrt{2}}\otimes\chi_{t,k}, k=1-3,
\end{aligned}
\tag{2}
$$

and for the other, in QD B,

$$
\begin{aligned}
{}_{0,1}^{2,0}X_{s}^{-1} &= \tilde{\varphi}_B(\mathbf{r}_h)\varphi_A(\mathbf{r}_{e,1})\varphi_A(\mathbf{r}_{e,2})\otimes\chi_s,\\
{}_{0,1}^{0,2}X_{s}^{-1} &= \tilde{\varphi}_B(\mathbf{r}_h)\varphi_B(\mathbf{r}_{e,1})\varphi_B(\mathbf{r}_{e,2})\otimes\chi_s,\\
{}_{0,1}^{1,1}X_{s}^{-1} &= \tilde{\varphi}_B(\mathbf{r}_h)\frac{\varphi_A(\mathbf{r}_{e,1})\varphi_B(\mathbf{r}_{e,2})+\varphi_A(\mathbf{r}_{e,2})\varphi_B(\mathbf{r}_{e,1})}{\sqrt{2}}\otimes\chi_s,\\
{}_{0,1}^{1,1}X_{t,k}^{-1} &= \tilde{\varphi}_B(\mathbf{r}_h)\frac{\varphi_A(\mathbf{r}_{e,1})\varphi_B(\mathbf{r}_{e,2})-\varphi_A(\mathbf{r}_{e,2})\varphi_B(\mathbf{r}_{e,1})}{\sqrt{2}}\otimes\chi_{t,k}, k=1-3.
\end{aligned}
\tag{3}
$$

The spin states of the trion have the form

$$
\begin{aligned}
\chi_s &= \tilde{\chi}_h\frac{\left|\uparrow_1,\downarrow_2\right\rangle-\left|\uparrow_2,\downarrow_1\right\rangle}{\sqrt{2}},\\
\chi_{t,1} &= \tilde{\chi}_h\frac{\left|\uparrow_1,\downarrow_2\right\rangle+\left|\uparrow_2,\downarrow_1\right\rangle}{\sqrt{2}},\\
\chi_{t,2} &= \tilde{\chi}_h\left|\uparrow_1,\uparrow_2\right\rangle, \chi_{t,3}=\tilde{\chi}_h\left|\downarrow_1,\downarrow_2\right\rangle,
\end{aligned}
\tag{4}
$$

where $\left|\uparrow(\downarrow)_k\right\rangle$ is the spin component of the *k*-th electron with spin projection $S_z = \pm 1/2$, and $\tilde{\chi}_h$ is the spin component of the heavy hole with spin projection $S_z = \pm 3/2$.

Since the hole wave function is localized in the central part of the QD due to its larger effective mass than that of the electron (see Section 4), we neglect its tunneling. In this case, the states from different groups do not mix with each other. Additionally, we do not consider the exchange effects that cause the splitting of triplet states and their interaction with singlet states. According to the results of Ref. [29], these effects have a negligible impact on the system's dynamics. In what follows, we will be interested in processes that involve the transposition of an electron between the QD and the occupation of the QD by two electrons, which eliminates the use of triplet states. Thus, we leave three singlet states in each group. A laser field with a frequency close to the exciton transition frequency will cause optical transitions between single-electron and singlet trion states if the spins of the electrons are in opposite directions. Let the spin projection of the electron in the charge qubit on the *z*-axis be +1/2. In this case, the laser should have a right-handed circular polarization of $\mathbf{e}_{-}$, so that the electron and hole of the resulting exciton have spin projections of -1/2 and +3/2, respectively, according to the selection rules [8]. Let's redefine the DQD states we are interested in by numbering them from 1 to 8:

$$\begin{aligned}
&|1\rangle = |\varphi_A\rangle|\uparrow\rangle, |2\rangle = |\varphi_B\rangle|\uparrow\rangle, \\
&|3\rangle = |{}^{2,0}_{1,0}X_s^{-1}\rangle, |4\rangle = |{}^{0,2}_{1,0}X_s^{-1}\rangle, |5\rangle = |{}^{1,1}_{1,0}X_s^{-1}\rangle, \\
&|6\rangle = |{}^{2,0}_{0,1}X_s^{-1}\rangle, |7\rangle = |{}^{0,2}_{0,1}X_s^{-1}\rangle, |8\rangle = |{}^{1,1}_{0,1}X_s^{-1}\rangle.
\end{aligned} \tag{5}$$

The Hamiltonian (1) in the tight-binding approximation takes the form

$$\begin{aligned}
H_{DQD} = &\sum_{k=1}^{8} \varepsilon_k |k\rangle\langle k| - V_{1,2}\left(|1\rangle\langle 2| + |2\rangle\langle 1|\right) + \\
&+ J\left(|3\rangle\langle 4| + |4\rangle\langle 3|\right) - V_{A,A}\left(|3\rangle\langle 5| + |5\rangle\langle 3|\right) - V_{B,A}\left(|4\rangle\langle 5| + |5\rangle\langle 4|\right) + \\
&+ J\left(|6\rangle\langle 7| + |7\rangle\langle 6|\right) - V_{A,B}\left(|6\rangle\langle 8| + |8\rangle\langle 6|\right) - V_{B,B}\left(|7\rangle\langle 8| + |8\rangle\langle 7|\right).
\end{aligned} \tag{6}$$

Here,

$$\begin{aligned}
&\varepsilon_1 = \varepsilon_A, \varepsilon_2 = \varepsilon_B, \\
&\varepsilon_3 = 2\varepsilon_A + \tilde{\varepsilon}_A + U_{2A,A}, \varepsilon_4 = 2\varepsilon_B + \tilde{\varepsilon}_A + U_{2B,A}, \varepsilon_5 = \varepsilon_A + \varepsilon_B + \tilde{\varepsilon}_A + U_{AB,A}, \\
&\varepsilon_6 = 2\varepsilon_A + \tilde{\varepsilon}_B + U_{2A,B}, \varepsilon_7 = 2\varepsilon_B + \tilde{\varepsilon}_B + U_{2B,B}, \varepsilon_8 = \varepsilon_A + \varepsilon_B + \tilde{\varepsilon}_B + U_{AB,B}
\end{aligned}$$

are the energies of the basis states, $V_{1,2} = \int d\mathbf{r}_e \varphi_A^*(\mathbf{r}_e) H(\mathbf{r}_e) \varphi_B(\mathbf{r}_e)$ is the energy (rate) of single-electron tunneling, and other parameters are expressed through the matrix elements of the Coulomb interaction

$$\begin{aligned}
U_{i,j,k,l}^{e-e} &= 2\iint d\mathbf{r}_{e,1} d\mathbf{r}_{e,2} \frac{\varphi_i^*(\mathbf{r}_{e,1})\varphi_j^*(\mathbf{r}_{e,2})\varphi_k(\mathbf{r}_{e,1})\varphi_l(\mathbf{r}_{e,2})}{|\mathbf{r}_{e,1} - \mathbf{r}_{e,2}|}, \\
U_{i,j,k,l}^{e-h} &= 2\iint d\mathbf{r}_h d\mathbf{r}_e \frac{\varphi_i^*(\mathbf{r}_e)\tilde{\varphi}_j^*(\mathbf{r}_h)\varphi_k(\mathbf{r}_e)\tilde{\varphi}_l(\mathbf{r}_h)}{|\mathbf{r}_h - \mathbf{r}_e|}.
\end{aligned} \tag{7}$$

For the Coulomb energies included in the energies of three-particle basis states, we obtain the following expressions:

$$\begin{aligned}
&U_{2A,A} = U_{A,A,A,A}^{e-e} - 2U_{A,A,A,A}^{e-h}, U_{2B,A} = U_{B,B,B,B}^{e-e} - 2U_{B,A,B,A}^{e-h}, \\
&U_{AB,A} = U_{A,B,A,B}^{e-e} + J - U_{A,A,A,A}^{e-h} - U_{B,A,B,A}^{e-h}, \\
&U_{2A,B} = U_{A,A,A,A}^{e-e} - 2U_{A,B,A,B}^{e-h}, U_{2B,B} = U_{B,B,B,B}^{e-e} - 2U_{B,B,B,B}^{e-h}, \\
&U_{AB,B} = U_{A,B,A,B}^{e-e} + J - U_{A,B,A,B}^{e-h} - U_{B,B,B,B}^{e-h}, J = U_{A,B,B,A}^{e-e}.
\end{aligned} \tag{8}$$

The non-diagonal matrix elements describing their tunnel hybridization have the form

$$\begin{aligned} V_{A,A} &= \sqrt{2}\left(V_{1,2} + U^{e-e}_{A,A,A,B} - U^{e-h}_{A,A,B,A}\right), V_{B,A} = \sqrt{2}\left(V_{1,2} + U^{e-e}_{B,A,B,B} - U^{e-h}_{A,A,B,A}\right), \\ V_{A,B} &= \sqrt{2}\left(V_{1,2} + U^{e-e}_{A,A,A,B} - U^{e-h}_{A,B,B,B}\right), V_{B,B} = \sqrt{2}\left(V_{1,2} + U^{e-e}_{B,A,B,B} - U^{e-h}_{A,B,B,B}\right). \end{aligned} \tag{9}$$

The multiplier $\sqrt{2}$ in expressions (9) indicates an increase in the rate of tunnel transitions between trion states due to the presence of two particles. To control the dynamics of the charge qubit by transitioning from the logical subspace of DQD to the auxiliary trion subspace, a laser field is applied to the structure in the form of two synchronously switched pulses:

$$\mathbf{E}_{las}(t) = \mathbf{E}_{las,1}(t)\cos\left(\omega_{las,1}t\right) + \mathbf{E}_{las,2}(t)\cos\left(\omega_{las,2}t\right). \tag{10}$$

This field generates direct transitions with the creation of an electron and a hole in the same QD, $|1\rangle \leftrightarrow |3\rangle, |1\rangle \leftrightarrow |8\rangle, |2\rangle \leftrightarrow |7\rangle, |2\rangle \leftrightarrow |5\rangle$, and indirect transitions with the creation of an electron and a hole in different QDs, $|1\rangle \leftrightarrow |5\rangle, |1\rangle \leftrightarrow |6\rangle, |2\rangle \leftrightarrow |4\rangle, |2\rangle \leftrightarrow |8\rangle$. The frequencies of these transitions are equal to the energy differences of the corresponding states, $\omega_{m,n} = \varepsilon_n - \varepsilon_m$ $(\hbar \equiv 1)$. The interaction of the DQD with the laser field is described by the Hamiltonian

$$\begin{aligned} H_{las} &= \sum_{k=1,2}\left[\Omega^{(k)}_{1,3}|1\rangle\langle 3| + \Omega^{(k)}_{1,8}|1\rangle\langle 8| + \Omega^{(k)}_{2,5}|2\rangle\langle 5| + \Omega^{(k)}_{2,7}|2\rangle\langle 7| + H.c.\right]\cos\left(\omega_{las,k}t\right) + \\ &+ \sum_{k=1,2}\left[\tilde{\Omega}^{(k)}_{1,5}|1\rangle\langle 5| + \tilde{\Omega}^{(k)}_{1,6}|1\rangle\langle 6| + \tilde{\Omega}^{(k)}_{2,4}|2\rangle\langle 4| + \tilde{\Omega}^{(k)}_{2,8}|2\rangle\langle 8| + H.c.\right]\cos\left(\omega_{las,k}t\right). \end{aligned} \tag{11}$$

Here, the optical Rabi frequencies for direct (indirect) transitions are introduced as $\Omega^{(k)}_{i,j}\left(\tilde{\Omega}^{(k)}_{i,j}\right) = \langle i|\mathbf{r}\mathbf{E}_k(t)|j\rangle$. Thus, the full Hamiltonian of a single-electron DQD in a laser field takes the form

$$H = H_{DQD} + H_{las}. \tag{12}$$

The state vector of the system is represented as a decomposition into basis vectors:

$$|\Psi(t)\rangle = \sum_{k=1}^{8} c_k(t)|k\rangle \exp\left(-i\varepsilon_k t\right), \tag{13}$$

where $c_k(t)$ are time-dependent probability amplitudes. The state vector satisfies the Schrödinger equation

$$i\frac{\partial}{\partial t}|\Psi(t)\rangle = H|\Psi(t)\rangle \tag{14}$$

with an initial condition $|\Psi(0)\rangle = \alpha_0|1\rangle + \beta_0|2\rangle$. Our goal is to obtain the desired final distribution of probability amplitudes $|\Psi(T_{end})\rangle = \alpha|1\rangle + \beta|2\rangle$ during the quantum operation at the moment $t = T_{end}$ when the pulse is turned off.

## 3. Quantum operations on a charge qubit using trion states

First, we will formulate the principle that underlies the proposed algorithms for single- and two-qubit gates. The idea is to synchronously transfer both components of the initial vector from the logical subspace to specific components in the trion subspace using resonant laser pulses. It is important to note that although both optical transitions induced by the lasers result in the generation of the same exciton, their frequencies differ by the difference in the Coulomb energies of the trion components. At the same time, in order to perform an amplitude rotation of the state vector in the X-Z plane of the Bloch sphere, these components must have similar energies for effective hybridization. In this case, the electron density is coherently redistributed between these states through tunneling. At a certain point in time, corresponding to the required transformation of the state vector, the laser pulses cause a reverse transition from the auxiliary (trion) subspace to the logical (electronic) subspace. As a result, the probability amplitudes of the trion components are transformed into the probability amplitudes of the logical states, which completes the algorithm. To perform a phase shift operation (rotation in the X-Y plane of the Bloch sphere), on the other hand, the auxiliary states must be isolated from each other, and the phase shift will depend on the difference in their Coulomb energies.

The described approach for X-Z rotation requires several conditions to be met on the system parameters. First, the tunneling coupling of the logical states in the absence of an exciton must be suppressed, i.e., $V_{1,2} << |\Delta|$, where $\Delta = \varepsilon_B - \varepsilon_A$. Second, only a selected pair of trion states (e.g., $|7\rangle$ and $|8\rangle$) must satisfy the tunneling resonance condition, $|\Delta + U_{2B,B} - U_{AB,B}| << V_{B,B}$, while the remaining three pairs must be blocked: $V_{A,B} << |-\Delta + U_{2A,B} - U_{AB,B}|$, $V_{A,A} << |-\Delta + U_{2A,A} - U_{AB,A}|$, $V_{B,A} << |\Delta + U_{2B,A} - U_{AB,A}|$. This also includes the prohibition of two-particle electron transfer between the QDs: $J << |2\Delta + U_{2B,A} - U_{2A,A}|, |2\Delta + U_{2B,B} - U_{2A,B}|$. As we will see, this condition is guaranteed to be fulfilled due to the smallness of the $J$ value. Thirdly, it is necessary to ensure the selectivity of optical transitions by choosing the laser frequencies $\omega_{las,1}$ and $\omega_{las,2}$ so that they correspond to the frequencies of only two direct electron-trion transitions $|1\rangle \leftrightarrow |8\rangle$ and $|2\rangle \leftrightarrow |7\rangle$. Thus, the resonance conditions must be met, $|\omega_{1,8} - \omega_{las,1}| << \Omega_{1,8}^{(1)}$ and $|\omega_{2,7} - \omega_{las,2}| << \Omega_{2,7}^{(2)}$, to ensure the rapid and selected excitation of the trion. The remaining transitions must be suppressed by large detunings of their frequencies from the laser frequencies, including the transitions between the selected states, but in the field of another laser, $\Omega_{1,8}^{(2)} << |\varepsilon_{1,8} - \omega_{las,2}|$ and $\Omega_{2,7}^{(1)} << |\varepsilon_{2,7} - \omega_{las,1}|$. Fourth, the optical transition rates must be

significantly larger than the tunneling rate, $V_{B,B} << \Omega^{(1)}_{1,8}, \Omega^{(2)}_{2,7}$, so that both processes can be considered independent. Fifth, they must be equal, $\Omega^{(1)}_{1,8} = \Omega^{(2)}_{2,7}$, so that the conversion of both components of the qubit state vector to auxiliary trion states and back is synchronized. The sixth condition, $\max\left(\gamma_{rel}, \gamma_{deph}\right) << V_{B,B}$, which reflects the smallness of the rates of dissipative processes – relaxation and dephasing – compared to the tunneling rate, ensures the coherence of the system during the operation.

If all the listed conditions are met, then equation (14) is equivalent to the system of equations for the probability amplitudes,

$$\begin{cases} i\dot{c}_1 = \Omega c_8 \\ i\dot{c}_2 = \Omega c_7 \\ i\dot{c}_7 = -Vc_8 + \Omega c_2 \\ i\dot{c}_8 = -Vc_7 + \Omega c_1 \end{cases}, \qquad (15)$$

where $\Omega = \Omega^{(1)}_{1,8} = \Omega^{(2)}_{2,7}$ and $V = V_{B,B}$. The fourth condition ensures the separation of optical and tunnel transitions. If the duration of the laser pulses is $T_{las} = \pi/2\Omega$, and the time of the tunnel interaction is $T_V = \theta/V$, then, since $T_{las} << T_V$, the evolution of the system is represented as a sequence of three stages. In the first step, an optical transition is performed from the logical subspace of the DQD to the auxiliary trion subspace. In the second step, the state vector is rotated by an angle $\theta$ in the trion basis. In the third step, a reverse optical transition is performed to the logical subspace:

$$\begin{cases} i\dot{c}_1 = \Omega c_8 \\ i\dot{c}_2 = \Omega c_7 \end{cases}, \; 0 \le t \le T_{las}; \; \begin{cases} i\dot{c}_7 = -Vc_8 \\ i\dot{c}_8 = -Vc_7 \end{cases}, \; T_{las} \le t \le T_{las} + T_V; \; \begin{cases} i\dot{c}_1 = \Omega c_8 \\ i\dot{c}_2 = \Omega c_7 \end{cases}, \; T_{las} + T_V \le t \le 2T_{las} + T_V. \qquad (16)$$

Each of the systems (16) describes Rabi oscillations:

$$\begin{pmatrix} c_m(t) \\ c_n(t) \end{pmatrix} = \begin{pmatrix} \cos(\Omega_R t) & -i\sin(\Omega_R t) \\ -i\sin(\Omega_R t) & \cos(\Omega_R t) \end{pmatrix} \begin{pmatrix} c_m(t_{init}) \\ c_n(t_{init}) \end{pmatrix}, \qquad (17)$$

where $\{m, n\}$ are pairs of states {1, 8}, {2, 7}, and {7, 8}, $\Omega_R = \Omega$ for the first and third stages, and $\Omega_R = V$ for the second stage, $t_{init}$ is the initial time for the given stage. The components of the DQD state vector are transformed as follows:

$$\begin{aligned} &|1\rangle \xrightarrow{1} -i|8\rangle \xrightarrow{2} -i\left(\cos\theta|8\rangle + i\sin\theta|7\rangle\right) \xrightarrow{3} (-i)^2\left(\cos\theta|1\rangle + i\sin\theta|2\rangle\right) \\ &|2\rangle \xrightarrow{1} -i|7\rangle \xrightarrow{2} -i\left(\cos\theta|7\rangle + i\sin\theta|8\rangle\right) \xrightarrow{3} (-i)^2\left(i\sin\theta|1\rangle + \cos\theta|2\rangle\right) \end{aligned}. \qquad (18)$$

Thus, at the final time $2T_{las} + T_V$, the qubit's state vector is rotated by an angle $\theta$, up to a phase factor $\exp(i\pi/2)$.

The phase operation is also implemented by exciting a pair of trion states that do not interact with each other. As these states, we can choose the states $|3\rangle$ and $|5\rangle$, whose energy difference is $\Delta + U_{AB,A} - U_{2A,A}$. According to the second condition, the tunneling coupling between them is blocked. The presence of the parameter $\Delta$ is related to the phase factor in the laboratory frame, which is absorbed when we switch to the rotating frame. Therefore, the accumulation of the relative phase $\delta\varphi(t) = \left(U_{AB,A} - U_{2A,A}\right)t$ is determined by the difference in the Coulomb energies of the trion states $|3\rangle$ and $|5\rangle$. Note that in the case of the rotation operation in expressions (18), this factor is absent due to the resonance of the auxiliary states. The transformation of probability amplitudes over time $T_\varphi$ can be represented as

$$\begin{aligned} &|1\rangle \xrightarrow{1} -i|3\rangle \xrightarrow{2} -i\exp\left(-iU_{2A,A}t\right)|3\rangle \xrightarrow{3} (-i)^2 \exp\left(-iU_{2A,A}T_\varphi\right)|1\rangle, \\ &|2\rangle \xrightarrow{1} -i|5\rangle \xrightarrow{2} -i\exp\left(-iU_{AB,A}t\right)|5\rangle \xrightarrow{3} (-i)^2 \exp\left(-iU_{AB,A}T_\varphi\right)|2\rangle. \end{aligned} \tag{19}$$

It is generally accepted that the main obstacle to creating a quantum register from charge qubits is the complexity of performing conditional operations on distant qubits. To date, such operations have been implemented on neighboring charge qubits [19, 20]. Theoretically, it is possible to use a scheme with sequential state exchange between neighboring qubits to transport the state of one of the qubits to another in the register. However, if the distance between them is large, this scheme appears unreliable and costly. Another approach, demonstrated on superconducting elements and trapped ions, uses an intermediate state of an auxiliary system (a microresonator mode or a vibrational mode of an ion chain) to organize an indirect interaction between distant qubits. The principle of operation of such a scheme is the selective excitation of a mode depending on the state of the controlling qubit and the participation of an excitation quantum (a photon or a phonon) in performing a given operation on target qubit. We have previously considered a method that allows this algorithm to be adapted to charge qubits. Its peculiarity is the use of transitions between the ground and excited states of the QD in the laser and resonator fields for conditional generation of a photon [39]. However, this scheme has not been realized experimentally. Below, we will show that the trion state can play the role of an auxiliary state when performing conditional operations and measuring a qubit.

Let us choose an optical microresonator or a waveguide [40] as an intermediate system, the mode frequency of which is close to the frequencies of exciton transitions of DQD. The task is to transfer a photon of the laser field to the mode of the MR for one of the components of the qubit state vector. Since the electron-trion transition in the QD B excites states for which the condition of tunnel resonance is fulfilled, it is not suitable for this purpose, since the state of the qubit changes in this case. On the other hand, the generation of a direct exciton in the QD A leads to the excitation of trion states that are isolated from each other. If we identify the single-electron component of the DQD $|B\rangle = |2\rangle$ with the logical "one" state of the qubit, then the absorption of a laser photon during the transition $|2\rangle|vac\rangle_c \to |5\rangle|vac\rangle_c$ and the generation of a photon in the MR mode during the transition $|5\rangle|vac\rangle_c \to |2\rangle|photon\rangle_c$ lead to the entanglement of the qubit and the MR quantum field:

$$\left(c_1|1\rangle + c_2|2\rangle\right)|vac\rangle_c \to c_1|1\rangle|vac\rangle_c + c_2|2\rangle|photon\rangle_c . \tag{20}$$

The emitted photon can control the evolution of another qubit that also interacts with the MR mode, and the detection of the photon indicates that the qubit is in a specific state (measurement).

In the next two sections, the parameters of Hamiltonian (6) will be calculated within the framework of the microscopic model of the two-dimensional DQD, and the dynamics of the qubit will be investigated, taking into account all the transitions presented in expression (11).

## 4. Calculation of qubit parameters on a quasi-two-dimensional lateral DQD

As a specific example, consider a charge qubit on a DQD formed in the process of controlled Stransky-Krastanov crystallization in the same layer of a GaAs/InGaAs heterostructure with an increased indium content. Typically, two-dimensional parabolic or Gaussian functions are used to describe the potential of a QD in the horizontal X-Y plane, while a one-dimensional step function is used in the vertical Z direction. We assume that the characteristic dimensions of the QD in the lateral X-Y plane are significantly larger than their dimensions along the vertical Z axis. This allows us to limit ourselves to solving the two-dimensional stationary Schrödinger equation for the wave functions of electrons and holes. This structure, unlike the vertical configuration of DQDs, where the QDs in different heterolayers are stacked one above the other, provides more opportunities for controlling their properties [4, 41]. Moreover, it better aligns with the concept of a quantum chip with a planar geometry.

The distribution of indium in the QD formation region in the X-Y plane is a stochastic quantity that fluctuates around the average value in the process of self-organization (crystallization) of Stransky-Krastanov, and is described by a Gaussian function. Let us set the potentials of the crystalline quasi-two-dimensional $In_xGa_{1-x}As$ DQD for electrons and holes as

$$U_{e(h)}(x,y) = U_{A,e(h)} \exp\left[-\left(\frac{x+0.5d_c}{r_{x,A}}\right)^2 - \left(\frac{y}{r_{y,A}}\right)^2\right] + U_{B,e(h)} \exp\left[-\left(\frac{x-0.5d_c}{r_{x,B}}\right)^2 - \left(\frac{y}{r_{y,B}}\right)^2\right], \tag{21}$$

where $r_{x,A(B)}$ and $r_{y,A(B)}$ are the characteristic sizes (radii) QD A(B) along the $x$ and $y$ axes, and $d_c$ is the distance between the QD centers. Following the authors of Refs. [27, 42, 43], we express the depth of the QD through the concentration of indium x = $X_{In,A(B)}$ at the center of QD A(B), the bandgap difference between InAs and GaAs $\Delta E_g$ = 1.11 eV, and their relative shift $\eta_e$ = 0.7 ($\eta_h$ = 0.3) for electrons (holes):

$$U_{A(B),e(h)} = -\eta_{e(h)} \Delta E_g X_{In,A(B)}. \tag{22}$$

The energies $\varepsilon_{e(h)} < 0$ of electrons (holes) localized in the DQD are measured from the bottom of the conduction band (the top of the valence band) in GaAs, where $U_{e(h)}(x,y) = 0$. The wave functions of the charge carriers satisfy the Schrödinger equation

$$-\frac{\hbar^2}{2m^*_{e(h)}}\left(\frac{\partial^2}{\partial x^2}+\frac{\partial^2}{\partial y^2}\right)\Psi_{e(h)}(x,y)+U_{e(h)}(x,y)\Psi_{e(h)}(x,y)=\varepsilon_{e(h)}\Psi_{e(h)}(x,y) \tag{23}$$

with the potential (21). We assume that the effective mass of the hole is isotropic and equal to $m^*_h$ = 0.6$m_e$. When choosing the potential parameters, it should be taken into account that the energy difference $\Delta$ of the electron states in different QDs, on the one hand, should reliably ensure the tunnel blockade, but on the other hand, be compensated by the difference in Coulomb energies. The parameters that can be varied are the depth of the QD (the concentration of indium), its size, and the distance between the QD centers.

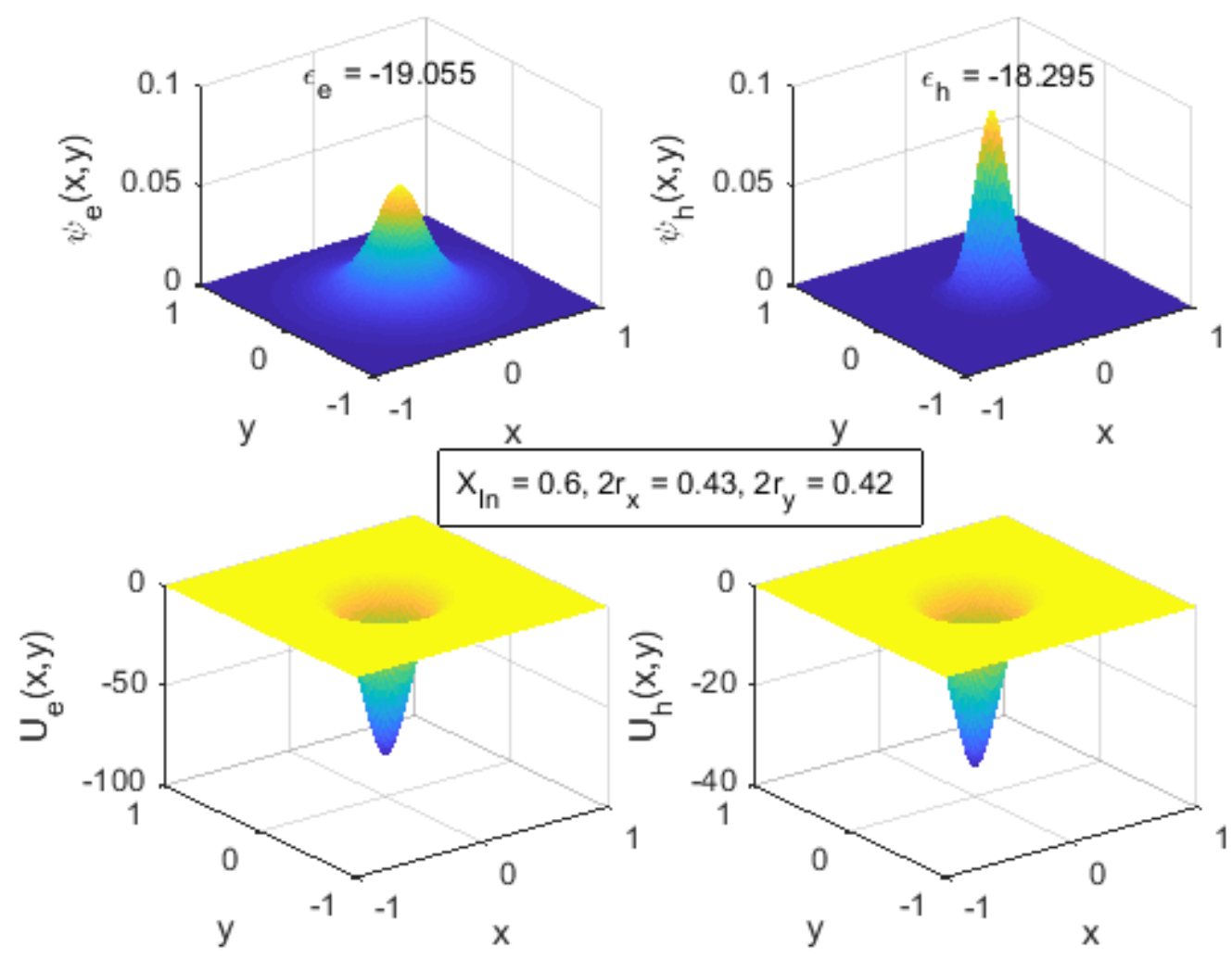

*Fig. 2. Wave functions, ground-state energies, and confinement potentials for electrons (left) and holes (right) in a single InGaAs QD. The potential parameters in effective atomic units are shown in the inset.*

Solving equation (23) by the finite-difference method, we find both the wave functions of the DQD and the wave functions of the isolated QD A and QD B (Fig. 2), which enter expressions (2), (3), and (7). The latter are obtained by keeping in expression (21) only one of the two components. As can be seen, the hole wave function is localized in the center of the QD more strongly than the electron wave function. Fig. 3 shows the dependence of the energies of the ground states of the electron and hole on the diameter of a single QD along the *x*-axis and its depth (indium concentration). These energies will be used to calculate the frequencies of exciton transitions in DQD and the detuning Δ.

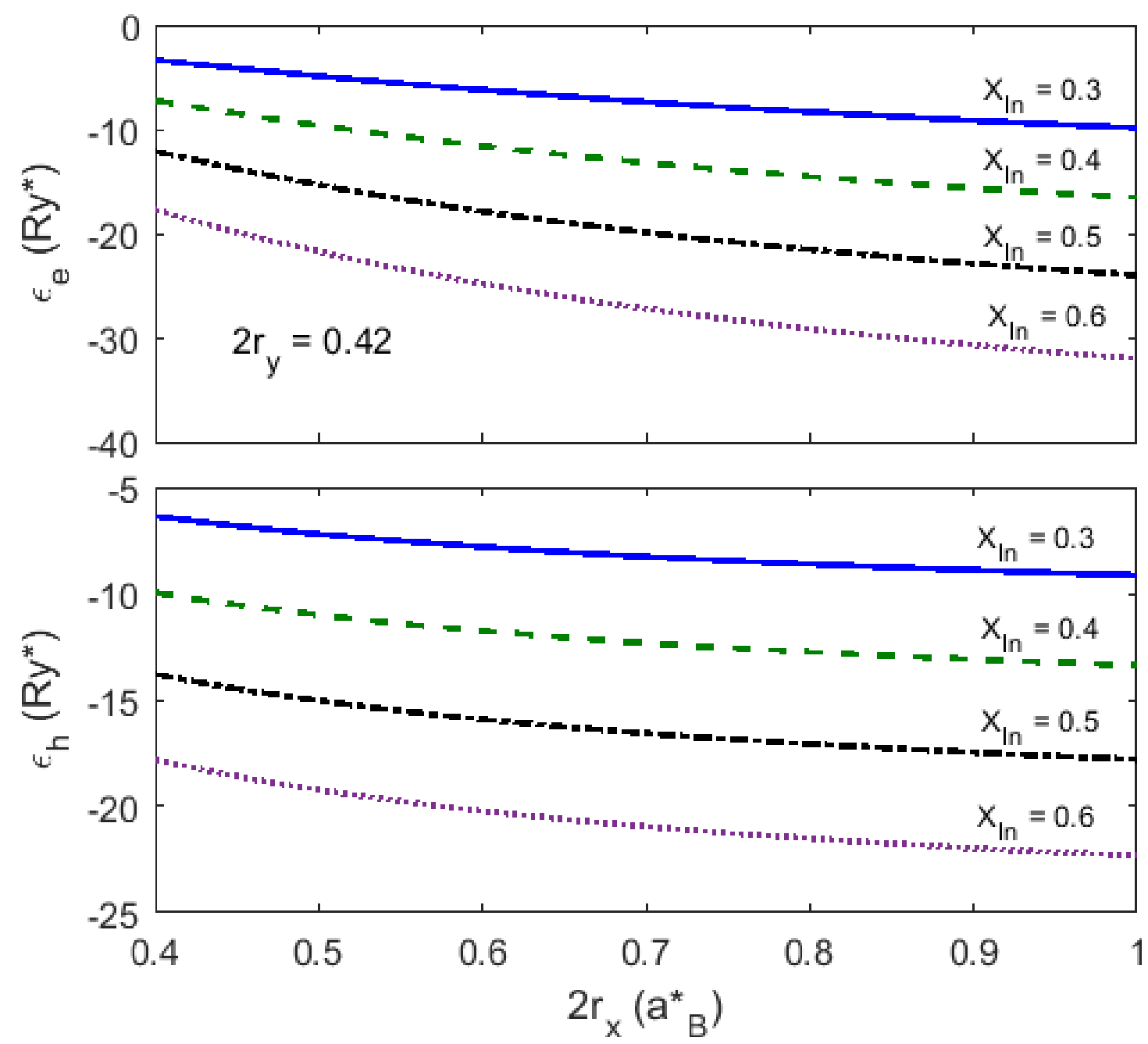

*Fig. 3. Dependences of the ground-state energies of the electron (top) and hole (bottom) in a single QD on its size along the x-axis, with a fixed size along the y-axis, for several values of the indium concentration (potential depth).*

The wave functions and energies of the DQD states are used to calculate the single-electron tunneling matrix element $V_{1,2}$. To do this, the indium concentration $X_{In,B}$ in QD B is varied relative to the fixed value $X_{In,A}$ in QD A. When these values match, the tunneling energy is equal to half the energy difference between the hybridized states. Figure 4 shows the dependence of $V_{1,2}$ on the thickness $b = d_c - r_{x,A} - r_{x,B}$ of the barrier separating the QDs, at $X_{In,A} = X_{In,B} = X_{In}$. As expected, it is approximated by an exponential function (a straight line on a logarithmic scale). Note that the tunneling matrix element for the hole is significantly smaller due to the localization of its wave function near the center of the KT, which supports the assumption of tunnel blockade in the valence band.

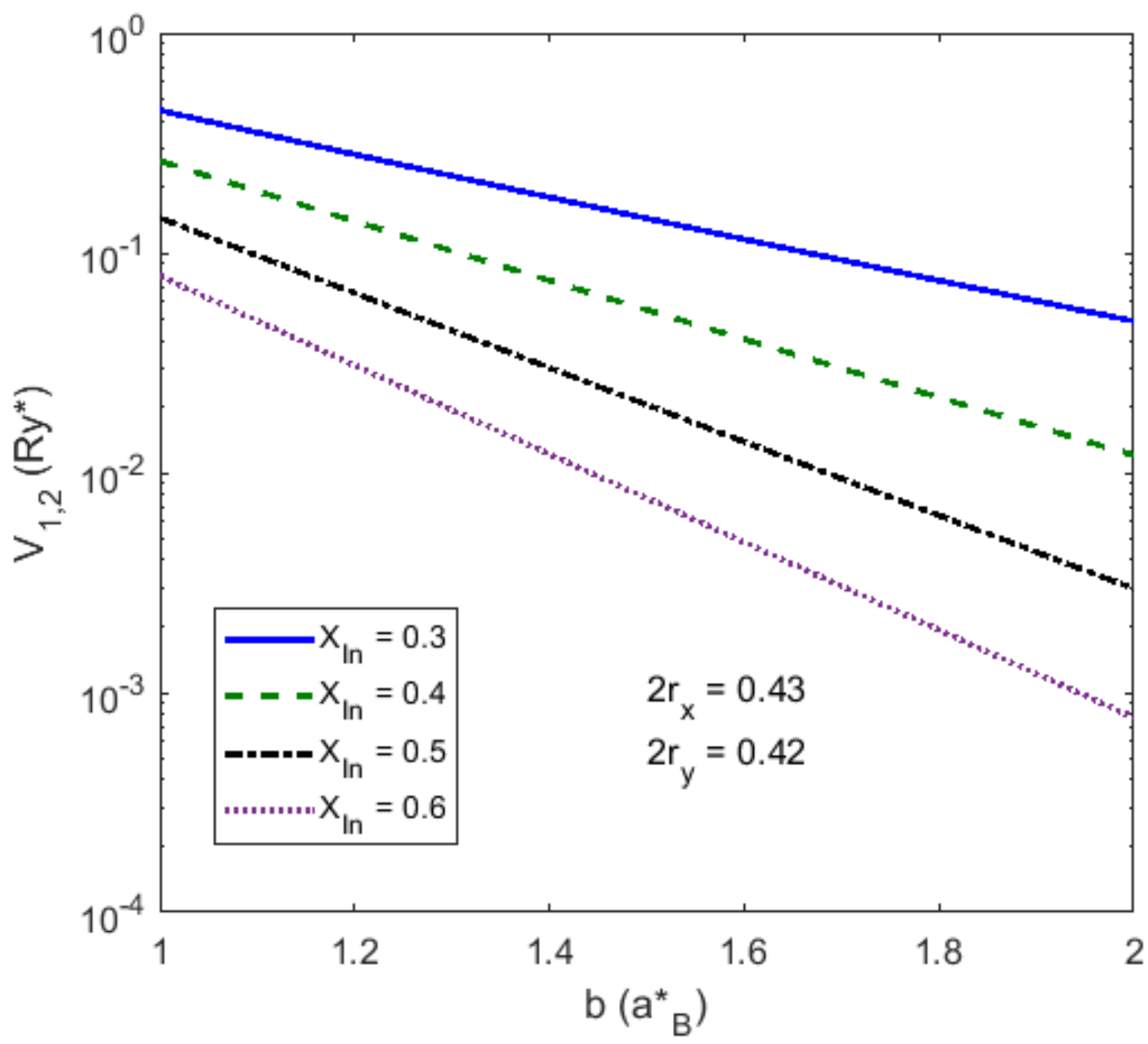

*Fig. 4. Dependences of the electron tunneling energy on the barrier thickness for several values of the indium concentration (potential depth of both QDs).*

The matrix elements of the Coulomb interaction, which appear in expressions (7) – (9), are divided into two groups, which are related to the localization of charge carriers in one or different QDs. In the first case, their dependence on the size of a given QD is of interest, and in the second case, their dependence on the distance between QDs. The difference in trion energies, which appears in the condition of resonant tunneling between "dressed" states, has the form

$$U_{2B,B} - U_{AB,B} = U^{e-e}_{B,B,B,B} - U^{e-h}_{B,B,B,B} - U^{e-e}_{A,B,A,B} + U^{e-h}_{A,B,A,B} - J \,. \quad (24)$$

First of all, we calculate the interaction energies of charge carriers in the same QD B as a function of its size and depth. The results of calculations (Fig. 5) indicate a stronger interaction between an electron and a hole than between two electrons. This is consistent with the data of experiments and numerical simulations [43] for vertical three-dimensional DQD. The value $U^{e-e}_{B,B,B,B} - U^{e-h}_{B,B,B,B}$ is several units of $Ry^*$, which corresponds to the interval of 5 – 15 meV.

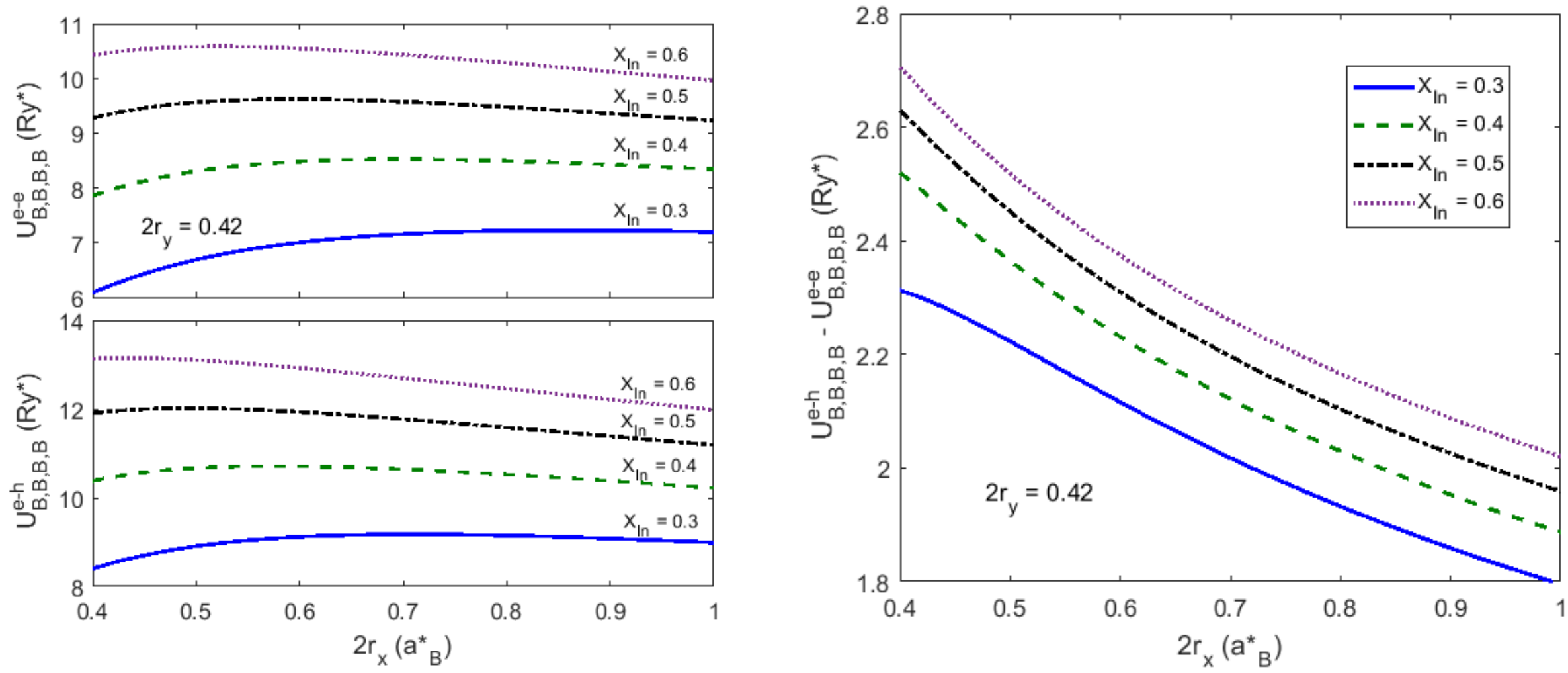

*Fig. 5. Dependences of the energies of electron-electron and electron-hole Coulomb interactions of particles (a) and their difference (b) in one QD on its size along the x-axis at a fixed size along the y-axis for several values of indium concentration (potential depth).*

The next order of magnitude is the interaction energy of charge carriers that are localized in different QDs. If the distance between the QDs is greater than their size, then these energies can be calculated with good accuracy using the point charge approximation $U_{A,B} \approx 2/d_c$ (Fig. 6a), and they are the same for electrons and holes. However, we need to obtain their difference, $U^{e-e}_{A,B,A,B} - U^{e-h}_{A,B,A,B}$, which can be calculated using expressions (7). As one can see, this difference is a few percent of the interaction energy difference in the same QD (Fig. 6b).

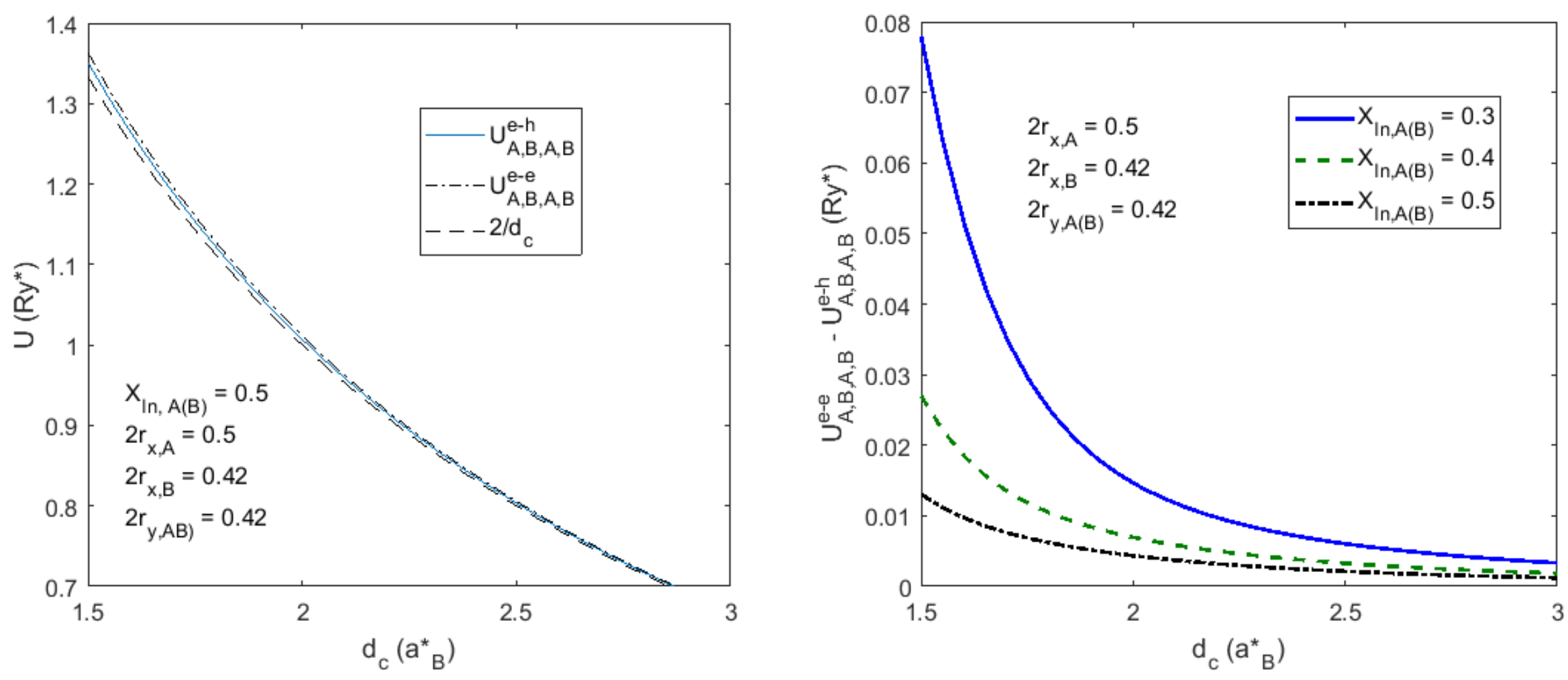

*Fig. 6. Dependences of the diagonal matrix elements of the Coulomb interaction (a) and their differences (b) for charge carriers in different QD on the distance between the centers of the QD for several values of indium concentrations.*

Finally, we present the data of calculations of the off-diagonal matrix elements $U^{e-h}_{A,B,B,B}$ and $U^{e-e}_{A,A,A,B}$, modifying the single-particle tunneling energy (9), and the exchange energy $J$. These quantities behave as $\exp(-\kappa d_c)$ and $\exp(-2\kappa d_c)$, respectively ($\kappa$ is a coefficient depending on the thickness and height of the barrier). In Fig. 7a, we present graphs of their difference, which is included in the first expression (9), as a function of the distance between the QD centers, and in Fig. 7b, we present the parameter $J$. At values of $d_c > 2$, their effect on the properties of the system is almost imperceptible, and, as indicated in Ref. [29], they can be ignored when considering the tunneling dynamics of an electron in a DQD.

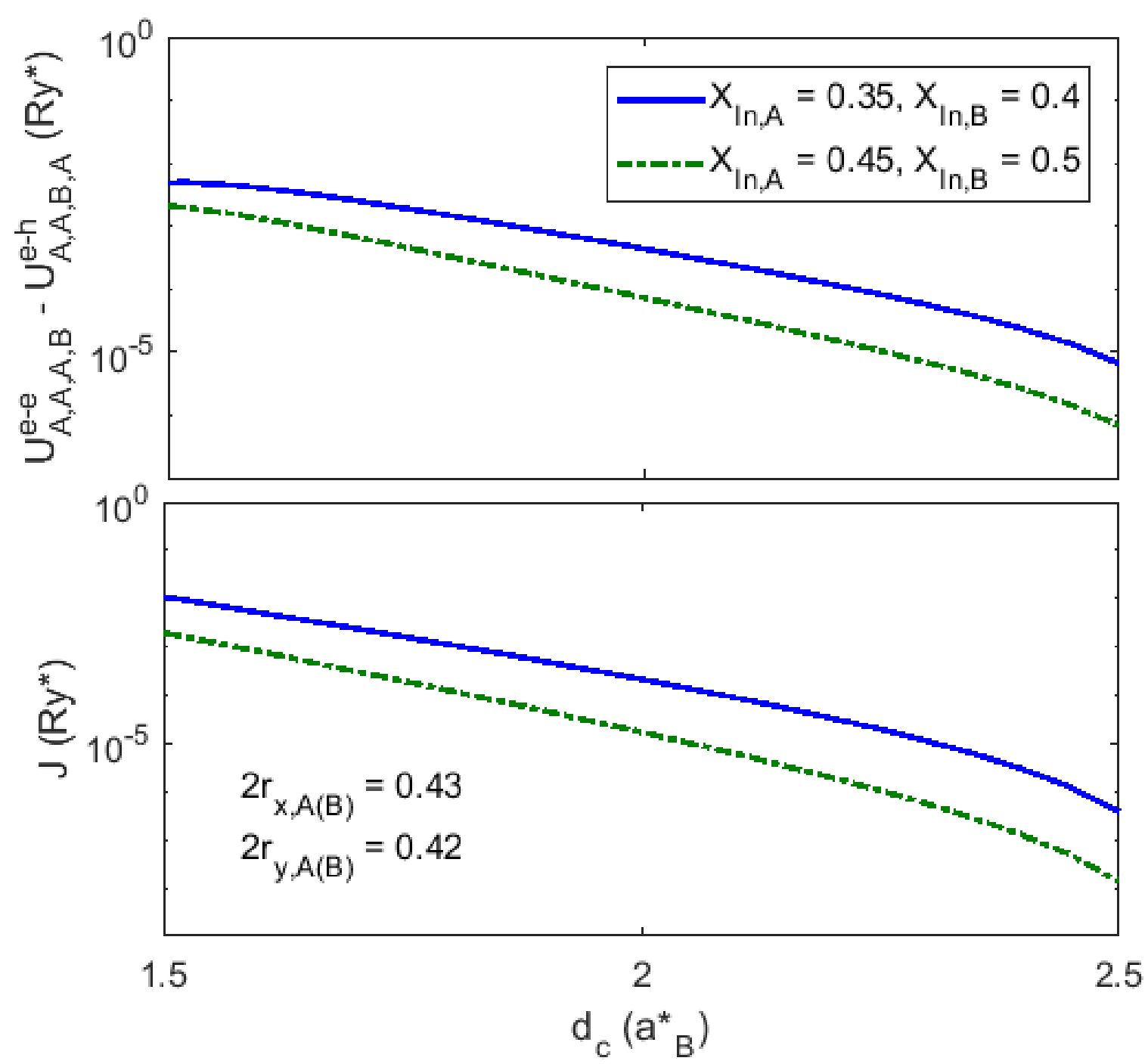

*Fig. 7. Dependences of the differences of the off-diagonal elements of the Coulomb interaction (top) and the exchange interaction energy (bottom) on the distance between the QDs for two sets of indium concentrations.*

Based on the data presented in Figs. 3-7, it is easy to calculate the frequencies of direct electron-trion transitions, which, given the choice of the energy reference point, are equal to

$$\begin{aligned}
\omega_{2,5} &= E_{g,A} + \varepsilon_A - U_{A,e} + \tilde{\varepsilon}_A - U_{A,h} + U_{AB,A}, \\
\omega_{1,3} &= E_{g,A} + \varepsilon_A - U_{A,e} + \tilde{\varepsilon}_A - U_{A,h} + U_{2A,A}, \\
\omega_{1,8} &= E_{g,B} + \varepsilon_B - U_{B,e} + \tilde{\varepsilon}_B - U_{B,h} + U_{AB,B}, \\
\omega_{2,7} &= E_{g,B} + \varepsilon_B - U_{B,e} + \tilde{\varepsilon}_B - U_{B,h} + U_{2B,B},
\end{aligned} \tag{25}$$

where the band gap of the $In_xGa_{1-x}As$ QD $i$ = A, B material depends on the relative concentration of indium x = $X_{In,i}$ and is given by the empirical expression in eV [44]:

$$E_{g,i} = 1.519 - 0.945x + 0.19x^2 + 0.019x^3 . \quad (26)$$

For the frequencies of indirect transitions, the band gap of one of the KT should be replaced by their average, $0.5(E_{g,A}+ E_{g,B})$.

A large group of parameters in the Hamiltonian forms a set of Rabi optical frequencies. If we ignore the spatial dependence of the amplitude of the laser field in the area of the DQD location, then the Rabi frequencies for electron-trion transitions can be expressed using the formula [8]

$$\Omega_{i,j}^{(k)} \left( \tilde{\Omega}_{i,j}^{(k)} \right) = \mu_{GaAs} E_{las,k} C_j \Phi_{i,j} , \quad (27)$$

where $\mu_{GaAs}$ is the matrix element of the dipole moment operator for the transition between the valence band and the conduction band in a GaAs crystal, $\Phi_{i,j} = \int \tilde{\varphi}_A(\mathbf{r}) \varphi_A(\mathbf{r}) d\mathbf{r}$ for pairs of indices $\{i, j\}$ = {1,3} and {2,5}, $\Phi_{i,j} = \int \tilde{\varphi}_B(\mathbf{r}) \varphi_B(\mathbf{r}) d\mathbf{r}$ for $\{i, j\}$ = {1,8} and {2,7}, $\Phi_{i,j} = \int \tilde{\varphi}_A(\mathbf{r}) \varphi_B(\mathbf{r}) d\mathbf{r}$ for $\{i, j\}$ = {1,5} and {2,4} and $\Phi_{i,j} = \int \tilde{\varphi}_B(\mathbf{r}) \varphi_A(\mathbf{r}) d\mathbf{r}$ for $\{i, j\}$ = {1,6} and {2,8}. The multiplier $C_j = 1/\sqrt{2}$ for $j$ = 5, 8, and $C_j = 1$ in other cases. Here, we have assumed that the DQD electron has a negligible effect on the single-particle wave functions of the resulting exciton, which are included in the integrand. This approximation works well in the regime of strong particle confinement in the QD potentials. For direct transitions, the integral in (27) is of the order of unity, since the electron and hole are created in the same QD, and in the case of an indirect transition, this integral depends on the overlap of the wave functions of carriers located in different QDs in the region of the barrier separating them and decreases exponentially with the growth of the barrier thickness. The value $\mu_{GaAs}$ is determined experimentally [45] or calculated using approximate formulas [46]. For three-dimensional crystalline GaAs, it is about $10^{-28}$ $C\times m$ or 40 Debay. We emphasize that the Rabi frequencies can be varied in a wide range by choosing the amplitudes of the laser fields.

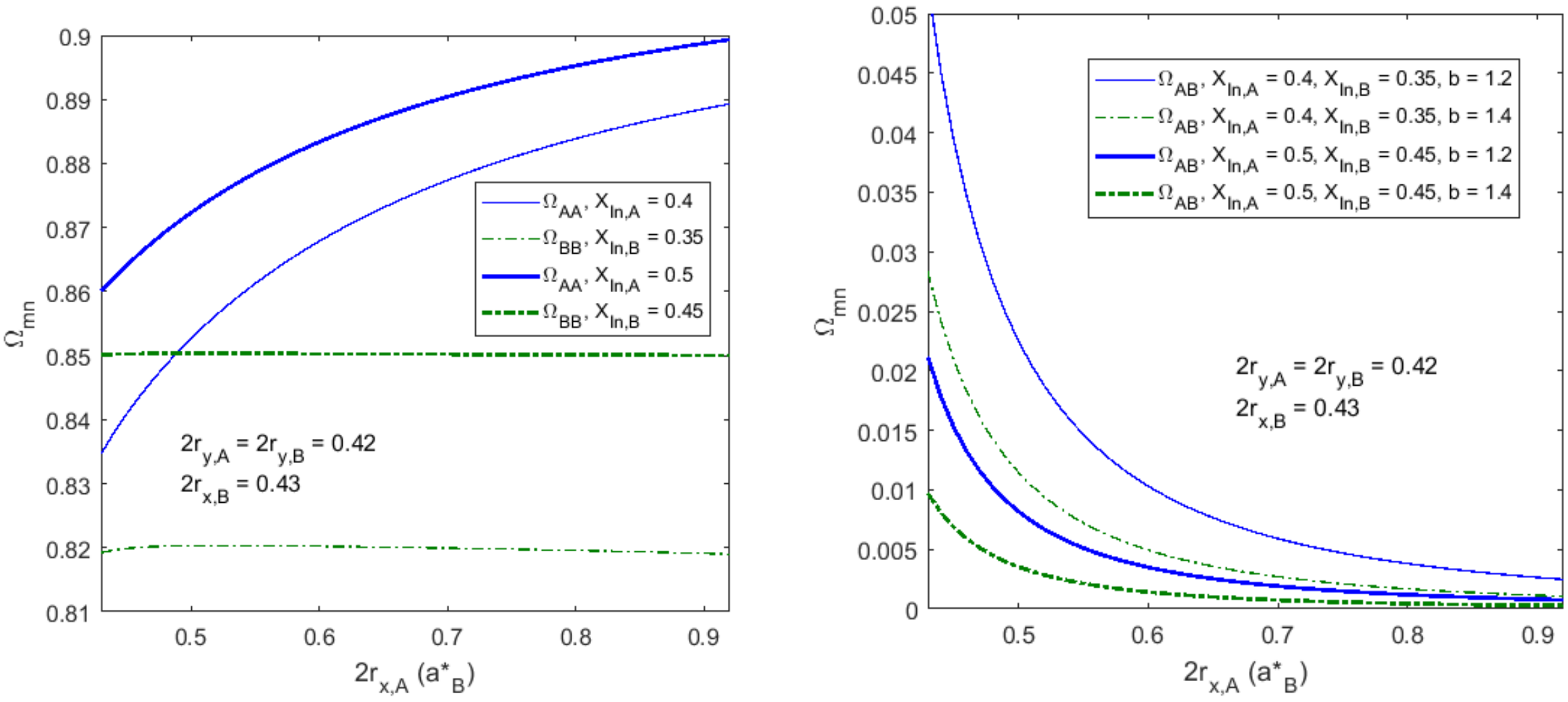

*Fig. 8. Dependences of the optical Rabi frequencies (in units of $\Omega_0$) for direct (a) and indirect (b) exciton transitions on the size of the QD A along the x-axis for two values of the indium concentrations and two values of the barrier thickness. The sizes of the QDB are fixed and are shown in the inset.*

Let us plot the dependence of the Rabi frequencies $\Omega_{mn} = \Omega_0 \int \tilde{\varphi}_m(\mathbf{r})\varphi_n(\mathbf{r})d\mathbf{r}$ ($m$, $n$ = A, B) in units of $\Omega_0 = \mu_{GaAs} E_{las,k}$ on the size of the QD A for several values of the depth of the QDs (the concentration of indium) and the thickness of the barrier between them (Fig. 8). It can be seen that as the linear size of the QD A increases along the x-axis, the Rabi frequency $\Omega_{AA}$ of the direct transition in QD A increases smoothly, while the Rabi frequency $\Omega_{BB}$ in the neighboring QD B remains constant. Both parameters increase for deeper QDs and are independent of the thickness of the barrier. On the contrary, the Rabi frequencies $\Omega_{AB}$ and $\Omega_{BA}$ of indirect transitions between QDs show a rapid decrease with increasing depth and width of the QD. This is due to the localization of the wave functions of charge carriers in the QD region and the decrease in the probability of their presence in the barrier region due to the deepening of the QD potential (increasing the energy difference between the level and the barrier top). It should be noted that the Rabi frequencies for indirect transitions are significantly lower than the Rabi frequencies for direct transitions.

The DQD parameters that satisfy the conditions formulated in Section 3 are selected as follows. First, we fix the size and depth of the QD B, where the exciton is excited, thus finding the carrier energies and the energy of their interaction in this QD. Then, based on the calculated data, we determine the size and depth of the second QD A, which satisfies the second condition (compensation of structural asymmetry and trion tunneling resonance). This choice also sets the exciton transition frequencies and the overlap integrals of the electron and hole wave functions in the DQD. Then, it is necessary to fix the barrier thickness so that the first condition (blocking the tunneling transfer of an electron without an exciton) and the sixth condition (coherence) are met. In addition, the difference in the Coulomb energies of particles in different QDs should strengthen condition (24). Finally, knowing the values of the overlap integrals for the Rabi frequencies, we set the amplitudes of the laser pulse components according to the third, fourth, and fifth conditions (selectivity and synchronism of resonant excitation of both transitions, and separation of tunneling and optical processes).

Let both QDs have the same chemical composition, i.e., the same indium concentration, $X_{In,A} = X_{In,B} = 0.5$ ($E_g = 182.73$). Thus, the depths of the QD potentials for the same type of particle are the same, $U_{e,A(B)} = -66.316$ and $U_{h,A(B)} = -27.75$. The diameters of QDs A and B along the x-axis, $r_{x,A} = 0.5$ and $r_{x,B} = 0.42$, differ by a small amount, while the diameters along the y-axis are chosen to be equal, $r_{y,A} = r_{x\backslash y,B} = 0.42$. Let's calculate the energies of the electron and the hole in given QDs: $\varepsilon_{e,A} = -15.43$, $\varepsilon_{e,B} = -12.83$, $\varepsilon_{h,A} = -15.07$, $\varepsilon_{h,B} = -14.09$. We use expressions (25) to calculate the frequencies of direct transitions, $\omega_{1,3} = 230.259$, $\omega_{1,8} = 236.353$, $\omega_{2,7} = 233.759$, $\omega_{2,5} = 232.712$. For a given set of parameters of the DQD potential, we find the difference in electronic energies, $\Delta = \varepsilon_{e,B} - \varepsilon_{e,A} = 2.6$, and the difference in Coulomb energies in QD B, $U^{e-e}_{B,B,B,B} - U^{e-h}_{B,B,B,B} = -2.594$. If we put $d_c = 2.46$ and, accordingly, $b = 2$, then we get $V_0 = V_{1,2} = 0.003$, $U^{e-e}_{A,B,A,B} - U^{e-h}_{A,B,A,B} = -0.001$, $U^{e-e}_{A,A,A,B} - U^{e-h}_{A,A,B,A} \simeq 10^{-6}$ and $J \simeq 10^{-8}$. Thus, the condition (24) is fulfilled with high accuracy, together with the condition for tunnel blocking for an electron. Using the data presented in Figure 8, we can calculate the overlap integrals for direct transitions, $\Omega_{AA} = 0.872\Omega_0$ and $\Omega_{BB} = 0.863\Omega_0$, and indirect transitions, $\Omega_{AB} = 0.0008\Omega_0$ and $\Omega_{BA} = 0.0009\Omega_0$. The selectivity and synchronism of excitation of direct transitions $|1\rangle \leftrightarrow |8\rangle$ and $|2\rangle \leftrightarrow |7\rangle$ require that their Rabi frequencies coincide and be significantly smaller than the detunings of their frequencies from the frequencies of other transitions. In our case, the minimum detuning is about 1 $Ry^*$, and we assume $\Omega^{(1)}_{1,8} = \Omega^{(2)}_{2,7} = 0.055$ to ensure the selectivity of resonant excitation and the separation of optical and tunneling processes. For crystalline GaAs, the Rabi frequency in effective atomic units is $\Omega_0 \approx 1.3\times10^{-5} E$, where the laser field amplitude $E$ is given in units of V/cm. Applying equation (27), we find the amplitudes of the laser fields corresponding to the Rabi frequencies: $E_{las,1} = 6.933$ kV/cm and $E_{las,2} = 9.804$ kV/cm. It is easy to calculate the remaining Rabi frequencies for these fields. In the next section, these parameters will be used to study the dynamics of charge carriers in DQD in an external laser field.

## 5. Modeling of the quantum NOT operation taking into account dissipative effects

Above, in Section 3, we have considered algorithms for performing single-qubit rotations around the X and Z axes of Bloch sphere in the coherent approximation and obtained analytical solutions of the Schrödinger equation corresponding to them. However, the presence of dissipative channels (relaxation and dephasing), as well as the excitation of non-resonant transitions involving states that are not involved in the algorithms, require a more rigorous analysis. The inclusion of processes leading to loss of coherence in the description of system dynamics is possible within the framework of an approach that uses the density matrix ρ and the Lindblad equation to model its evolution,

$$\frac{d\rho}{dt} = -i\left[H,\rho\right] + L_{rel}\left(\rho\right) + L_{deph}\left(\rho\right). \tag{28}$$

The exciton relaxation, which irreversibly converts the trionic state into the electronic DQD state, is introduced in the Markov approximation using the operator

$$\begin{aligned} L_{rel}\left(\rho\right) = \gamma_d \left[ D\left(|1\rangle\langle 3|\right) + D\left(|1\rangle\langle 8|\right) + D\left(|2\rangle\langle 5|\right) + D\left(|2\rangle\langle 7|\right) \right] + \\ + \gamma_i \left[ D\left(|1\rangle\langle 5|\right) + D\left(|1\rangle\langle 6|\right) + D\left(|2\rangle\langle 4|\right) + D\left(|2\rangle\langle 8|\right) \right], \end{aligned} \tag{29}$$

and the qubit dephasing, which causes a stochastic failure of the relative phase of its components in both the logical and trion subspaces, is described by the operator

$$\begin{aligned} L_{deph}\left(\rho\right) = \gamma_q D\left(|1\rangle\langle 1| - |2\rangle\langle 2|\right) + \gamma_t \left[ D\left(|3\rangle\langle 3| - |5\rangle\langle 5|\right) + D\left(|4\rangle\langle 4| - |5\rangle\langle 5|\right) + \right. \\ \left. + D\left(|7\rangle\langle 7| - |8\rangle\langle 8|\right) + D\left(|6\rangle\langle 6| - |8\rangle\langle 8|\right) \right]. \end{aligned} \tag{30}$$

Here, $\gamma_d$ ($\gamma_i$) are the relaxation rates of the trion due to direct (indirect) transitions, and $\gamma_q(\gamma_t)$ is the dephasing rate of the electron (trion) state of the DQD. The expression $D\left(O\right) = O\rho O^{\dagger} - \left\{O^{\dagger}O,\rho\right\}/2$ models the decay of a quantity corresponding to the $O$ operator (the curly brackets denote the anticommutator). In all subsequent calculations, the rates of non-coherent processes are included as independent phenomenological parameters.

The amplitudes of the laser pulses are set as smooth functions of time on intervals corresponding to the on/off processes. They are represented by tangential dependencies

$$E_{las,k}\left(t\right)=E_{las}^{(k)}\left[\tanh\left(\frac{t-t_0^{(k)}}{\tau_0^{(k)}}\right)-\tanh\left(\frac{t-t_0^{(k)}-T_0^{(k)}}{\tau_0^{(k)}}\right)+\tanh\left(\frac{t-\tilde{t}_0^{(k)}}{\tau_0^{(k)}}\right)-\tanh\left(\frac{t-\tilde{t}_0^{(k)}-\tilde{T}_0^{(k)}}{\tau_0^{(k)}}\right)\right],\quad (31)$$

where $t_0^{(k)}$ – the beginning of stage 1, $\tilde{t}_0^{(k)}=t_0^{(k)}+T_{las}+T_V$ – the beginning of stage 3, $T_0^{(k)}=\tilde{T}_0^{(k)}=T_{las}$ – the duration, and $\tau_0^{(k)}$ – the on/off time for the *k*-th pulse. We assume that the parameters $t_0^{(k)}$ may differ for the two lasers, violating their synchronization: $t_0^{(2)}=t_0^{(1)}+\Delta T$. Figure 9 shows the time dependence of the populations of the system states (the diagonal elements $\rho_{n,n}$ of the density matrix) illustrating the implementation of the NOT operation (inversion of the qubit states) for the set of parameters defined in Section 4. The direct trion relaxation rate $\gamma_d \sim 10^{-4} – 10^{-5}$ *Ry** in (29) corresponds to the times $\tau_d \sim 10^{-9} – 10^{-8}$ s observed in experiments. The increase in the trion relaxation time by an order of magnitude compared to the exciton ($10^{-10} – 10^{-9}$ s) is due to the destructive interference of two equivalent collapse channels caused by the presence of two electrons in the QD instead of one. Indirect relaxation is even slower due to the weak overlap of the electron and hole wave functions in different QDs, and has little effect on the system's dynamics. We assume that the rates of direct relaxation and dephasing are identical. The results are consistent with the analytical solution (18) in Section 3, which was obtained without considering dissipation. At intervals where laser pulses are applied, a weak perturbation of the harmonic Rabi dependences can be observed, which is associated with non-resonant excitation of other transitions. The process of tunnel transfer between states 7 and 8 is accompanied by a slight increase in the populations of states 1 and 2, which is caused by the decay of the trion. The operation time is approximately 500 1/*Ry**, or about 50 picoseconds, which is comparable to similar times for electrically controlled charge qubits and exciton qubits. The second stage is the most time-consuming, as the tunneling transfer rate must be significantly lower than the optical excitation rate. However, the resulting value allows for the demonstration of single-qubit rotations in DQD with acceptable accuracy at the current technological level.

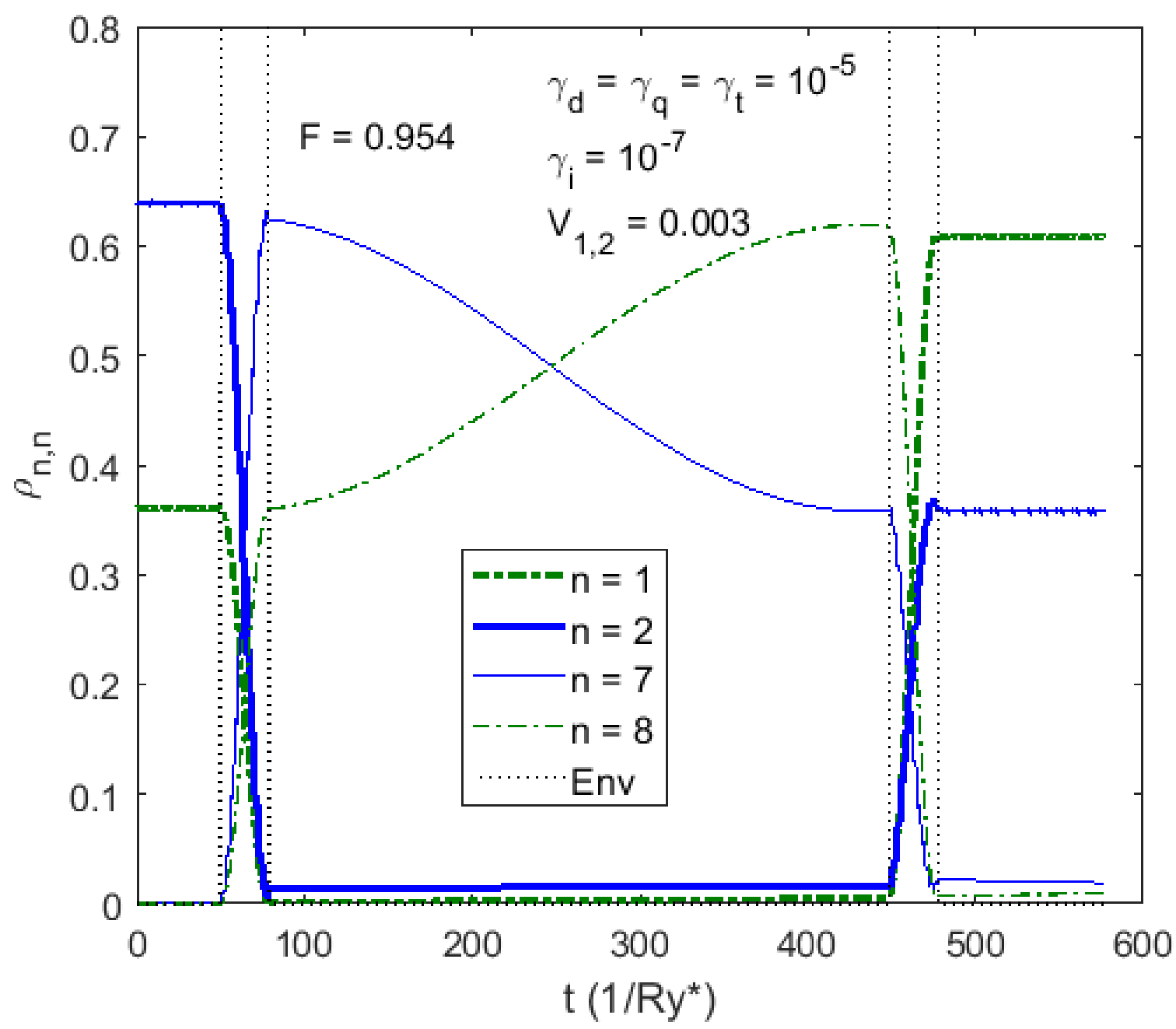

*Fig. 9. Dependences of the population of qubit states on time for the NOT operation ($\alpha$ = 0.6, $\beta$ = 0.8). The intervals corresponding to the generation and annihilation of an exciton by a laser field (stages 1 and 3) are indicated by dotted lines. Here,* $T_0^{(1)} = T_0^{(2)} = T_{las}$, *$\Delta T$ = 0 and* $\tau_0^{(1)} = \tau_0^{(2)} = 0.01T_{las}$.

Usually, the function that characterizes the quality of performing a single-qubit operation $U$ is the fidelity $F$. If the initial state $\rho_{in}$ and the final state $\rho_{end} = U\rho_{in}$ of the system for a perfectly executed operation are pure, then it is expressed by the formula $F(t) = tr\left[\rho_{end}\rho(t)\right]$, where $\rho(t)$ is the solution of equation (28). It is of interest to study the fidelity from the amplitude of the laser field, as well as from the rates of tunneling, decay, and dephasing. The dependence of $F(T_V + 2T_{las})$ on the parameter $E_{las}^{(k)} / E_0$, where $E_0$ is the amplitude of the $k$-th pulse defined in Section 4, for two trion decay rates is shown in Figure 10. It is possible to distinguish three intervals with curve behavior, related to the specifics of the system dynamics. For weak fields ($0 < E_{las}^{(k)} / E_0 < 0.5$), when the optical excitation rate is less than or comparable to the tunneling rate, the fidelity is low. This is due to the overlapping of optical and tunneling processes, which leads to a more complex four-level scheme than the two-level Rabi model [47]. In this case, the probability of complete population transfer between the two trion states is low. For fields with amplitudes $0.5 \le E_{las}^{(k)} / E_0 \le 1.5$, $F$ reaches its maximum value and stabilizes. This occurs due to the separation of the aforementioned processes. A further increase in the amplitude is accompanied by a decrease in $F$, which takes the form of a quasi-periodic dependence. This indicates the involvement of other states in the system's dynamics, which are caused by non-resonant processes [48]. Therefore, there is an interval of field amplitude values where, at low decay rates, high values of $F \geq 0.95$ are observed.

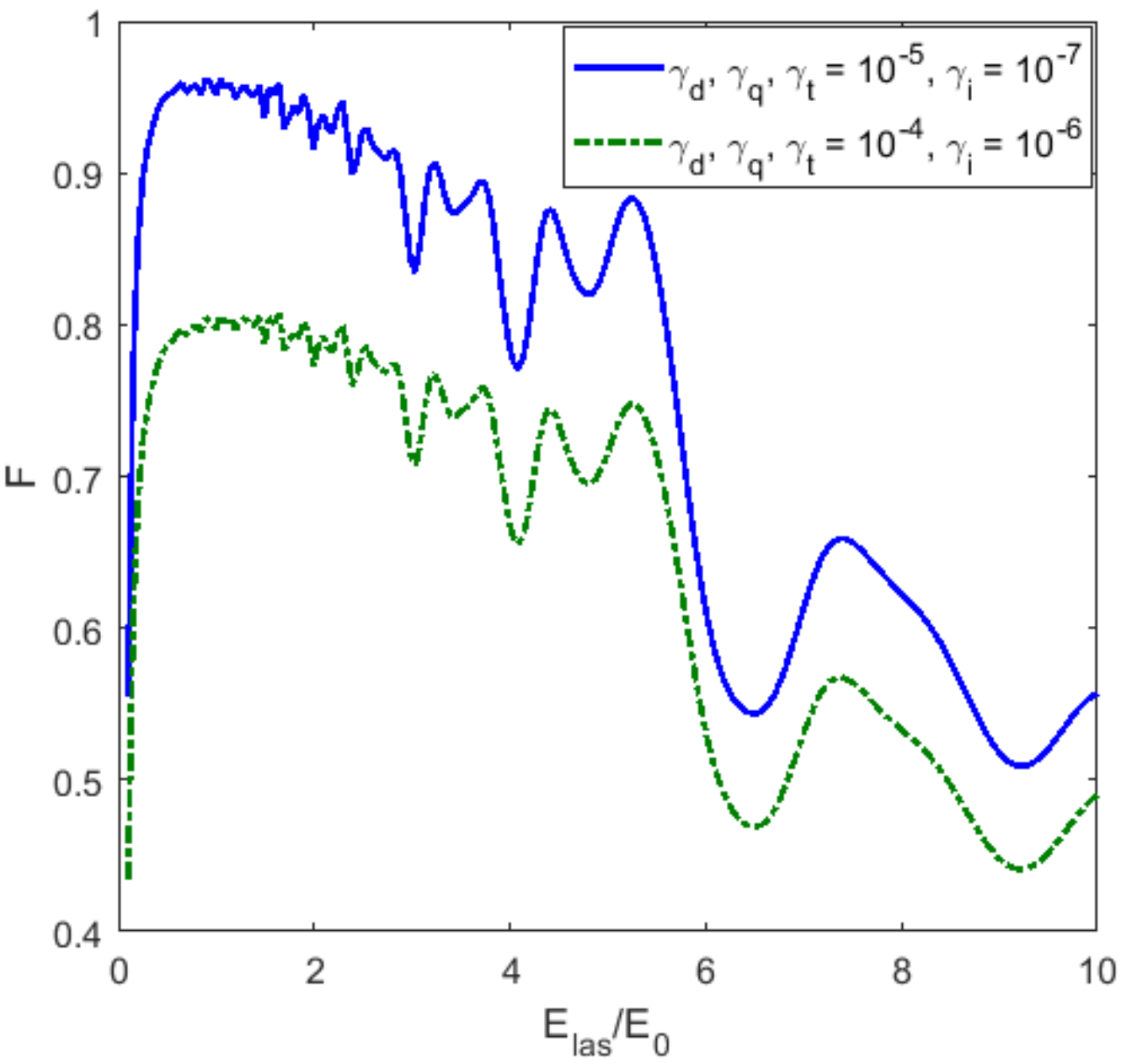

*Fig. 10. Fidelity as a function of the laser field amplitude for low and moderate dissipation cases.*

Suppression of unwanted transitions (increasing selectivity) at fixed frequencies requires weakening the laser field, but this also slows down the addressed optical transitions. Although reducing the Rabi frequencies by an order of magnitude preserves the coherence of the system's evolution for the specified dissipation rates, increasing the mutual modulation of the tunneling and optical processes significantly reduces the fidelity. Therefore, it is necessary to simultaneously reduce the tunneling rate, but not to the extent that non-coherent mechanisms become dominant. In our case, the parameter $V_{1,2}$ is varied by changing the thickness $b$ of the barrier between the QDs. It does not affect the basic parameters associated with the characteristics of individual QDs, affecting only the difference in the energies of the Coulomb interaction of carriers in different QDs. Therefore, we assume that other quantities do not change when $V_{1,2}$ is changed. The dependence of $F$ on $V_{1,2}$ is shown in Fig. 11. As can be seen, the slowing down of tunneling at $V_{1,2} \leq 0.5V_0$ is accompanied by a rapid decrease in $F$, caused by an increase in the influence of dissipative effects on the trionic state. An increase in $V_{1,2}$ above $3V_0$ also leads to a decrease in the fidelity, which is associated with an increase in the mutual influence of tunnel and optical processes. However, for a mode with moderate dissipation, there is a slight increase in $F$ over the range $V_0 \leq V_{1,2} \leq 2V_0$ due to a reduction in the duration of the second stage.

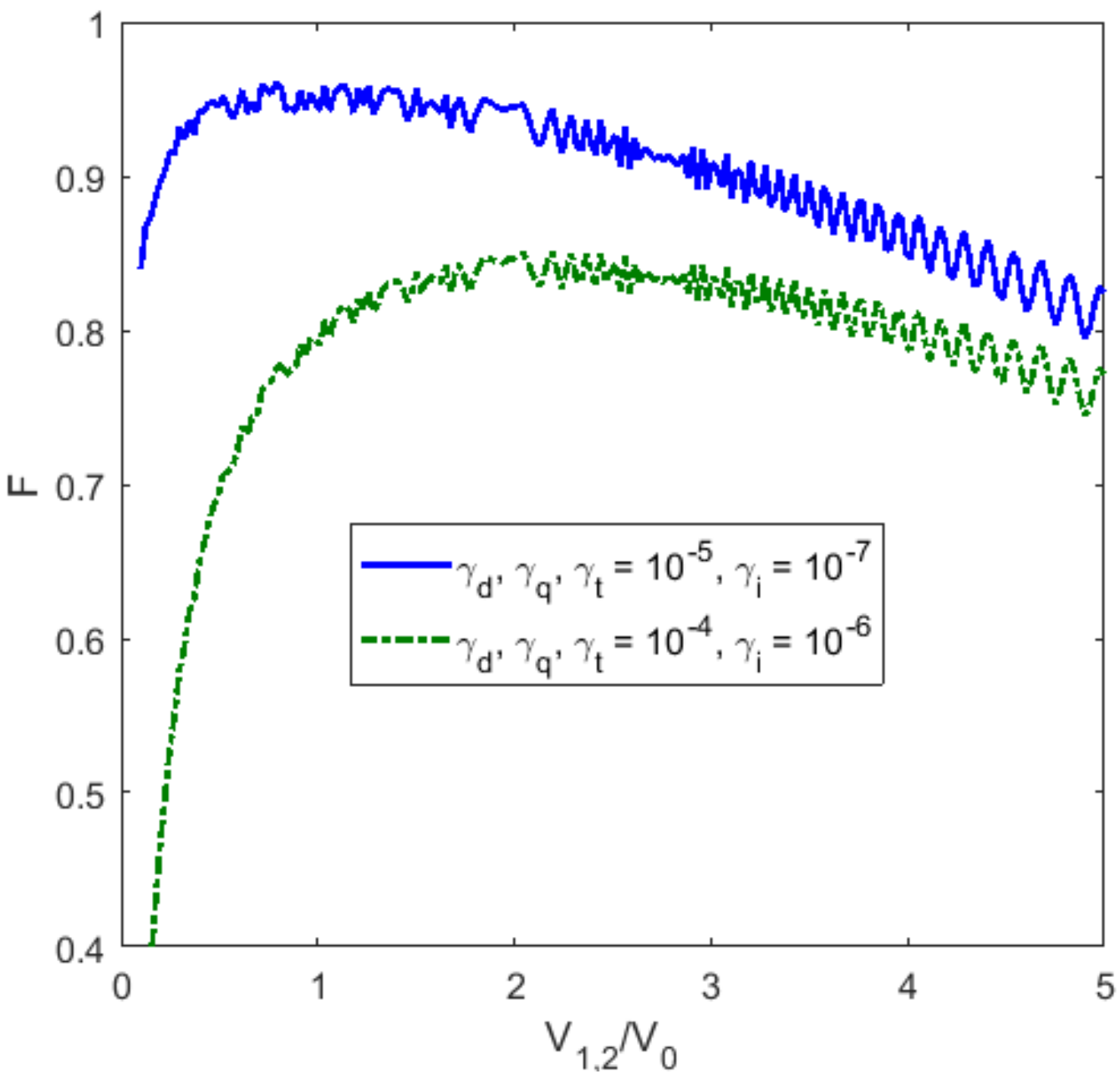

*Fig. 11. Fidelity as a function of tunneling energy for low and moderate dissipation cases.*

Finally, we will study the effect of asynchronism on the fidelity by introducing a time delay $\Delta T$ between laser pulse sequences 1 and 2. In a three-level resonant excitation scheme [47], even a small delay leads to a sharp decrease in $F$ due to a violation of the Rabi frequency balance. However, as shown in Figure 12, in a scheme with separation of tunneling and optical processes, high values of $F$ are maintained over a wide range of delays. The smooth decrease in $F$ begins only when $|\Delta T| > T_{las}$, when the pulses stop overlapping. This behavior is explained by the small change in the trion component excited by the earlier pulse due to tunneling during the generation of the other trion component by the delayed pulse. Therefore, the proposed algorithm has a certain tolerance for controlling the pulse durations.

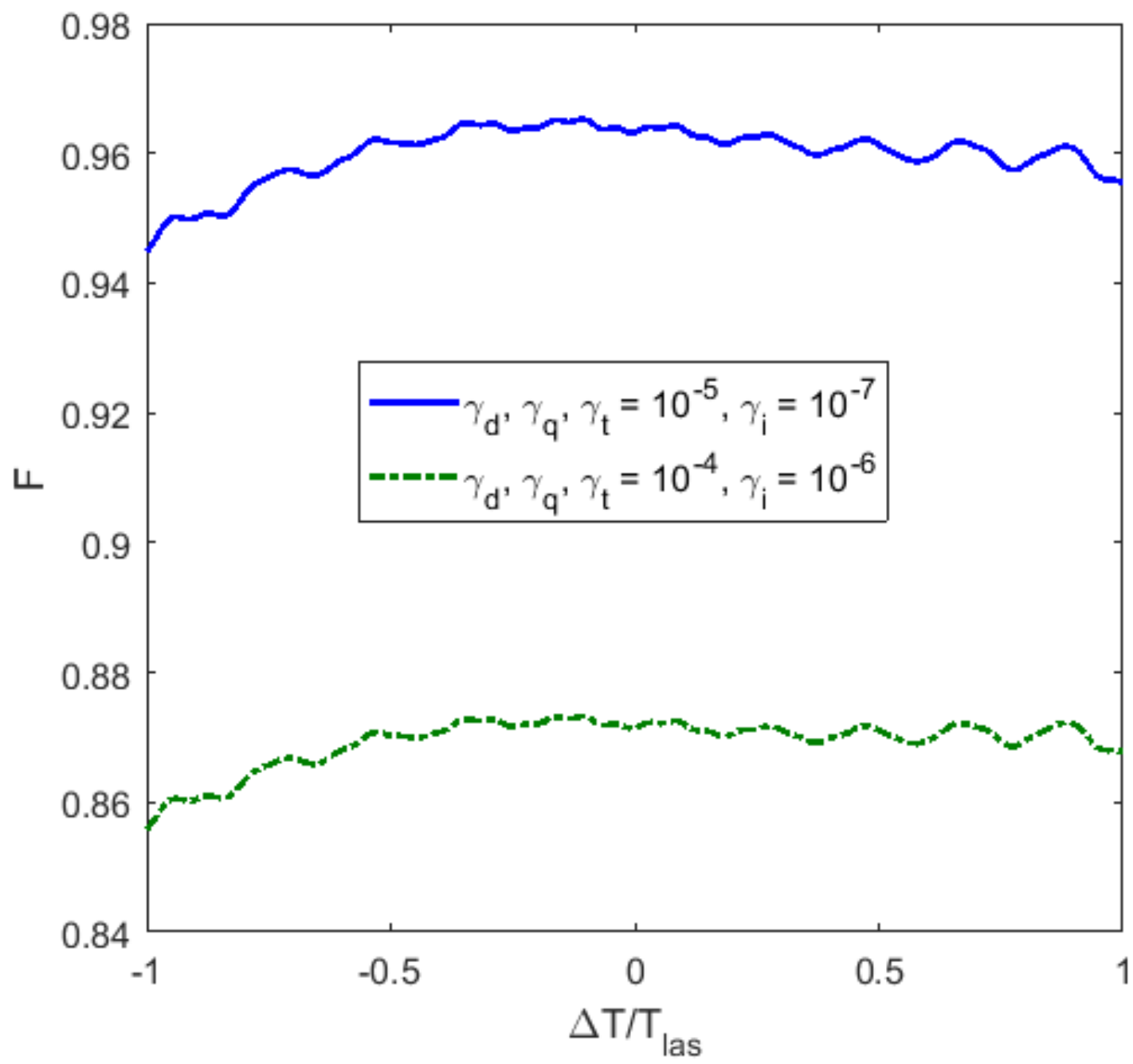

*Fig. 12. Fidelity as a function of relative laser pulse delay for low and moderate dissipation cases.*

We add that the dependencies of the fidelity on the dissipation rates demonstrate the standard behavior reflecting the exponential decay of $F$ with the growth of $\gamma_{d(i,q,t)}$. In our work, we do not provide separate graphs for $F(\gamma)$, limiting ourselves to the data for two sets of relaxation and dephasing rates in Fig. 10 – 12.

## 5. Discussion

The physical properties of trions in QDs are markedly different from similar complexes formed in three-dimensional crystals or layered heterostructures [49]. Experimental studies have shown that the spectra of singly and doubly charged excitons in DQD demonstrate a strong dependence on an external electric field applied along its axis [23, 29]. In particular, at a certain value of the field amplitude, an anti-crossing of trion states localized in different QDs is observed, which indicates their hybridization due to resonant tunneling of an electron or a hole. Since the condition of resonance of electronic (hole) states is not satisfied for the given field in the single-particle DQD, the excitation of an exciton in such a system leads to a rapid (~ 1 ps) activation of the tunnel interaction. As follows from the analysis of the spectrograms, the simple tight-binding model takes into account all the key points related to the behavior of the trion in the DQD. The microscopic modeling of spectra and calculation of energies and wave functions of trions were performed using the variational method [27] and the direct solution of the Schrödinger equation using finite-difference methods [28, 43]. The results obtained in the tight-binding approximation are quantitatively consistent with numerical results and experimental data, provided that the fitting parameters are chosen correctly. Since the potential profiles of the valence band and conduction band differ, the resonances of electronic and hole states are observed at different values of the external field. This means that effective hybridization is possible only for one type of charge carrier. The authors of Ref. [29] emphasize that two phenomenological parameters are sufficient to describe a trion in a DQD: the single-particle tunneling energy and the difference in Coulomb energies between two trion states corresponding to the localization of mobile carriers in the same or different QDs. Additionally, the use of the tight-binding approximation and the Hubbard model to represent the Hamiltonian in terms of electronic and trion states is in line with the roadmap for solid-state qubit research outlined in Ref. [50]. The dependence of photoluminescence on an external electric field allows the type of particle participating in hybridization to be determined by the size of the anti-crossing, the tunneling energies and the time of the trion decay to be calculated. The main features of the spectra should be considered a red shift (~ 5 – 6 meV) of the trion line relative to the neutral exciton line in QD and an increase in the lifetime compared to the exciton, which is explained by the specifics of the Coulomb correlations between particles. We note that for the compact quasi-two-dimensional QDs considered in our work, this shift is slightly higher due to the increased localization of particles compared to three-dimensional QDs.

An important feature (and complexity) of the described mechanism for compensating structural asymmetry using the Coulomb interaction of three particles is the need to satisfy the second condition in Section 3 with very high accuracy. In the main order (with an accuracy of $10^{-2}$ $Ry^*$), the asymmetry is eliminated due to the negative contribution in equation (24), which is related to the interaction of two electrons and a hole in the DQD. This is achieved by selecting the size and chemical composition of the QD, taking into account technological fluctuations. However, the tunneling energy in our model is quite small (on the order of $10^{-3}$ $Ry^*$) due to the need to separate the tunneling and optical processes. Therefore, in order to maintain resonant tunneling transport, it is necessary to control the compensating energy with an accuracy of at least $10^{-4}$ $Ry^*$, which is unlikely to be achieved during the manufacturing process of the DQD. Let us assume that there is a weak constant electric field along the DQD axis, which corrects the compensation mechanism, increasing the accuracy to the required level [51]. It can be created by one or more gates. This field should be considered as a passive auxiliary tool, which does not directly control the evolution of the qubit, as in other gate-based schemes. By adding it to the chip architecture, the resonance of the trion states can be achieved with the required accuracy.

We also mention other approaches to implementing quantum computations on QDs based on qubit state control mechanisms that use an auxiliary state. In Refs. [47, 51-55], single-qubit operations via electron transpositions in DQDs are proposed using resonant transitions that connect logical spatially separated *s*-states localized in different QDs through an excited state that is delocalized over both QDs. In contrast to the scheme under consideration, here the tunneling energy is markedly larger than the Rabi optical frequency. This leads to hybridization of the electronic *p*-orbitals of individual QDs and the formation of a doublet of DQD molecular states. To excite the transitions, classical laser radiation and/or a quantum single-photon field of a microresonator with frequencies corresponding to the electronic transition between the *s*- and *p*-states of the QDs are used. The resonant frequencies of the transitions between the ground and excited states are shifted by the value of $\pm V$ with respect to the frequencies of isolated QDs. The choice of control field frequencies determines the type of qubit evolution. Tuning the field to resonance with the transition frequency to one of the doublet states leads to a three-level dynamics, which provides a fast inversion, but does not allow for an arbitrary amplitude transformation of the qubit. On the contrary, a large detuning from the resonance generates an effective two-level Raman dynamics and allows to carry out this operation with high fidelity of reproduction, but with a much slower speed than in the resonant scheme. In our case, two components of the trion subspace are hybridized, $|7\rangle$ and $|8\rangle$. At the same time, the trion multiplet contains four more states, and therefore eight direct transition frequencies, which are located in a fairly narrow spectral interval. In this regard, the application of the Raman scheme involving trions requires further study. To date, there are no reports of successful experimental implementation of optical control of a charge qubit.

The authors of theoretical works [56 – 58], considering the vacuum state and the indirect exciton in the asymmetric DQD as logical states of a qubit, organize the interaction between them through the direct exciton, which is transformed into the indirect one due to the resonant tunneling of an electron between the QDs. This scheme requires the inclusion of a control gate that regulates the shift of the electronic states. The spatial separation of the electron and the hole contributes to the increase in the exciton lifetime. In Ref. [59], a model of a qubit on a triple QD is proposed, whose logical states are represented by two indirect excitons with a hole in the central QD and electrons in the left QD or right QD. Here, single-qubit rotations also require tuning the energy levels of the electrons using an external time-dependent gate field. Conditional two-qubit operations are based on the excitation of a four-particle complex in a controlled qubit. The trend of using multi-particle states for information encoding has been practically implemented in the concept of the so-called hybrid qubit [60]. It is an optically controlled DQD with three electrons, has a high sensitivity to electric fields and a long coherence time. It has demonstrated single-qubit operations with a Rabi frequency of more than 100 MHz and $F > 0.93$. It is worth noting that the trion states used in our scheme as replicas of the qubit states maintain high coherence during the operation time. Thus, it is advisable to use them as auxiliary ones in small time intervals, but not for long-term storage of information. Finally, in structures with more complex geometry, consisting of a large number of tunnel-coupled QDs, trions can act as components of delocalized fermion complexes [61].

## 6. Conclusions

This work is devoted to the study of issues related to the use of a negatively charged trion for implementing quantum operations on a charge qubit. In our proposed scheme, single-qubit rotations are performed using resonant optical transitions between single-electron (logical) and trion (auxiliary) states of the qubit. The asymmetry of the DQD confinement potential creates an energy shift in the logical states, which blocks electron tunneling and freezes the evolution of the qubit. The energy difference of the Coulomb interaction of the trion particles compensates for this shift, stimulating the tunneling connection between them. The resulting (transformed by tunneling) trion state is then translated back into the logical subspace of the DQD, completing the algorithm. The requirements for the structure and control field necessary for the successful implementation of the proposed algorithm have been formulated.

Using the example of a two-dimensional GaAs/InGaAs DQD with lateral tunnel coupling, the energies of the electron and hole, the matrix elements of the tunnel and Coulomb interactions, and the optical Rabi frequencies were calculated. The dependencies of these parameters on the QD dimensions, chemical composition, and distance between the QDs were analyzed. The evolution of the qubit was described using a tight-binding model that takes into account dissipative effects. It was shown that the fidelity of the inversion operation reaches high values ($F > 0.95$) over a certain range of parameters. At the same time, the system is highly resistant to laser pulse asynchronism.

**Acknowledgements:** The work is carried out within the state assignment of NRC “Kurchatov institute”.

**Author contributions:** I am the sole author of this work and no contributions from others were involved in its creation.

**Data availability:** No datasets were generated or analysed during the current study.

**Conflict of interest**: The author confirms that there are no known competing financial interests or personal relationships associated with this publication for this work that could have influenced its outcome.